\documentclass[a4paper,10pt]{article}
\pdfoutput=1
\usepackage[pagewise]{lineno} %%-------- For gettline line numbers ----------  %% debug debug %%
\usepackage{amsfonts} %%------ AMS Fonts package ------------------
\usepackage{amsmath} %%------- Graphics package -------------------
\usepackage{amssymb} %%------- AMS symbol package?-----------------
\usepackage[pdftex]{graphicx}   %% --- Graphics package -------------------
\usepackage{color}  %%-------- Color package ----------------------
\usepackage{wrapfig} %%------- Wrapping text around a figure ------
\usepackage{epsfig} %%-------- EPS support? -----------------------
\usepackage{wasysym} %%------- ??? --------------------------------
\usepackage{pifont} %% ------- ??? --------------------------------
\usepackage{mathrsfs} %%------ Special math fonts -----------------
\usepackage{array} %%--------- Table formatting aids---------------
\usepackage[pdftex,x11names,svgnames]{xcolor} %%-------y
\usepackage[pdftex]{hyperref} %%------ Hyperlink citations and references--
\usepackage{algorithm} %%----- For typesetting pseudo-code---------
\usepackage{algorithmic} %%--- For typesetting pseudo-code---------
\usepackage{cancel} % --------- For cancelling expressions---------
\usepackage{tikz} %------------ For inline graph & diagrams -------
\usetikzlibrary{positioning} %- Tikz library ----------------------
\usetikzlibrary{arrows} %------ Arrows ----------------------------
\usepackage{verbatim} 
\usetikzlibrary{arrows,shapes} 
\newcommand{\TT}{\scriptscriptstyle} %%----------- Smallest size---
\newcommand{\vb}[1]{{\boldsymbol {#1}}} %%-------- For vector face-

\newcommand{\ud}{\mathop{}\!\mathrm{d}} %%--------differential-----
\definecolor{brown}{rgb}{0.65,0.325,0} %%--------- Brown color ----
\definecolor{purple}{rgb}{1,0.5,0.75} %%---------- Blue color -----
\begin{document}
\title{Classical 5D fields generated by a uniformly accelerated point source}

\author{ I. Aharonovich$^{a}$
         and
         L. P. Horwitz$^{abc}$ \\
         \\
         $^{a}$ Bar-Ilan University, Department of Physics, Ramat Gan, Israel.  \\
         $^{b}$ Tel-Aviv University, School of Physics, Ramat Aviv, Israel. \\
         $^{c}$ College of Judea and Samaria, Ariel, Israel. \\
       }

%------------------------------------------------------------------
\maketitle
%------------------------------------------------------------------

\begin{abstract}
Gauge fields associated with the manifestly covariant dynamics of particles in $(3,1)$
spacetime are five-dimensional. In this paper we explore the old problem of 
fields generated by a source undergoing hyperbolic motion in this framework. 
The 5D fields are computed numerically
using absolute time $\tau$-retarded Green-functions, and qualitatively compared with Maxwell fields generated
by the same motion. 
We find that although the zero mode of all fields coincides with the corresponding Maxwell problem,
the non-zero mode should affect, through the Lorentz force, the observed motion of test 
particles.
\end{abstract}

%%\tableofcontents
%%\listoffigures

\section{Introduction}
    \label{sec:introduction}
    %%
%% File   :  UAP_Introduction
%% Purpose:  The introduction section of the uniform-acceleration paper.
%%
The problem of Maxwell $(3,1)$ electromagnetic fields generated by a uniformly accelerated point source 
is an age old problem starting as early as 1909 by Max Born \cite{born_1909_1}, followed by a plethora
of papers and books\footnote{A far from inclusive list includes 
                              \cite{bondi_gold_1955,born_1909_1,Fulton1960,ginzburg_1970,Rohrlich1990,boulware_1980,Parrott:1993ig,parrott_1997,Harpaz:1998wd,vallisneri_2000}.}.
                               
The problem appears deceptively simple, and yet, it has raised many arguments and discussions
on the very nature of radiation and inertial motion.

Locally, a uniformly accelerating source is essentially a notion from Newtonian Mechanics, 
mostly associated with a point mass in a static
constant gravitational field, or an electric charge in similarly static and constant electric field.
A covariant description of this motion corresponds to a hyperbola in a spacetime of $1+1$ dimensions,
a worldline of the parametrized form
\begin{align}
    \label{eq:intro_hyperbolic_worldline_1}
    z(\tau) & = z_0 + \dfrac{1}{g} \cosh (g\tau)
    \\
    \label{eq:intro_hyperbolic_worldline_2}
    t(\tau) & = t_0 + \dfrac{1}{g} \sinh (g\tau)
\end{align}
where $g$ is the Newtonian acceleration in the local frame (LF) $g = d^2z/dt^2 |_{\text{LF}}$, 
and $\tau$ denotes \emph{proper time}.

The main issues that have been debated are:
\begin{enumerate}
 \item Uniform acceleration does not cause radiation reaction. It can be shown in the (mass renormalized) Lorentz-Dirac equation
       \cite{dirac_1938,Fulton1960,boulware_1980,Poisson:1999tv} 
       \begin{align}
            \label{eq:intro_lorentz_dirac}
            m \ddot{z}^{\mu} & = \Gamma_{\text{ext}}^{\mu} + \Gamma_{\text{rr}}^{\mu}
                               = \Gamma_{\text{ext}}^{\mu} 
                                 +
                                 \dfrac{2 e^2}{3 c^2}   
                                 \left[
                                    \dddot{z}^{\mu} 
                                    +
                                    \dfrac{1}{c^2} \left( \ddot{z}^{\nu} \ddot{z}_{\nu}\right) \dot{z}^{\mu}
                                 \right]
       \end{align}
       that the \emph{radiation reaction} force term $\Gamma_{\text{rr}}^{\mu}$ \emph{vanishes}
       for the motion described by \eqref{eq:intro_hyperbolic_worldline_1} and \eqref{eq:intro_hyperbolic_worldline_2}.
       This apparent contradiction of the conservation of energy
       can be resolved \cite{boulware_1980,Parrott:1993ig,Harpaz:1998wd,vallisneri_2000},
       either by limiting the time interval in which the particle is 
       accelerated \cite{ginzburg_1970,Parrott:1993ig,parrott_1997}, 
       or, alternatively, draining the apparent inexhaustible self-energy of the particle's field \cite{Fulton1960}.
       Locally, uniform-acceleration has an event horizon at $x=t$, and cannot be influenced by events outside
       this cone. However, this means that locally, the charge is unaffected by its own loss of energy due to radiation.
       
       In the limit of unbounded accelerated motion, the balance of mechanical and electromagnetic energy
       becomes a balance of infinities, and is not well defined \cite{vallisneri_2000}.
       
 \item In the special inertial frame in which the source is \emph{momentarily at rest},  
       the \emph{magnetic field vanishes for all space}. This might lead to an apparent contradiction 
       as the \emph{Poynting vector is zero everywhere} in this frame, \emph{suggesting there is a lack of radiation}, 
       opposing the well known formula for overall radiation generated by an accelerating point source
       \cite{Rohrlich1990,Jackson1995}
       \begin{align*}
            R & = \dfrac{2 e^2 \ddot{z}_{\mu} \ddot{z}^{\mu}}{3c^3}
       \end{align*}
              
       This apparent contradiction is easily resolved (e.g. \cite{Fulton1960,vallisneri_2000})
       by noting the fact that the magnetic field value $\vb{H}(\vb{x},t_{0})$ on a given $t = t_{0}$ 3D surface
       depends \emph{on the entire history of the source particle} (prior to $t_{0}$).
       However, the transfer of radiated energy follows along the light cone 
       (via the conservation equation for the energy-momentum tensor $\theta^{\mu \nu}$), 
       and not along space-like surfaces.
       
 \item In the setting of general relativity, a somewhat more uncertain question remains: 
       How does radiation transpire in the presence of a constant static background gravitational field?
       For instance, \cite{boulware_1980,Parrott:1993ig,vallisneri_2000} claim that radiation takes place
       in the case of a fixed (supported) detector observing a point charge which undergoes geodesic (freefall) motion, whereas
       \cite{singal_1997,Harpaz:1998wd} claim the opposite.
       The difference amongst claims has its roots in the very notion of radiation, of whether it is a property
       of space \cite{Harpaz:1998wd} or of the observer \cite{Parrott:1993ig,vallisneri_2000}.
       \footnote{
           In this paper, however, we shall not address this issue altogether, as it is dedicated to the 
           relativistic dynamics without regard to gravity and the equivalence principle.
       }       
\end{enumerate}

In this paper, we examine the problem in a different framework, namely, the 
\emph{off-shell electrodynamics} arising in the manifestly covariant relativistic dynamics of 
Stueckelberg \cite{Stueckelberg1941,Stueckelberg1942} and
Horwitz and Piron \cite{HorPir1973,SaadHorArsh1989}.

In this framework, a \emph{universal time $\tau$} is defined, parameterizing the motion 
of all \emph{events}, which can be viewed schematically as $4D$ spacetime particles. 
Therefore, the resultant electromagnetic fields, which may depend on a spacetime point $x^{\mu}$
\emph{as well as $\tau$},  generally obey a \emph{five dimensional} linear wave-equation,  in either $(4,1)$ or $(3,2)$ symmetry.
Both the additional dimension and the requirement for causality in $\tau$, lead to a different support structure of the 
corresponding Green-function. The domain of dependence of a field $a^{\alpha}$ at a 5D observation point $(x^{\mu},\tau)$
is now the \emph{entire retarded $\tau$} history of the source inside the corresponding 5D past 
hyper-cone\footnote{The exact form of the hypercone depends on the choice of $(4,1)$ or $(3,2)$ metric of the field.}.
Our conclusion is that there is indeed 
radiation in this configuration; even though the zero mode of the 5D 
fields coincides with the standard Maxwell fields, the Lorentz force due 
to the non-zero modes can have an effect on test particles.

% % % The resultant field generated by such motion differs substantially from the Maxwell field, even though that asymptotically, 
% % % in $\tau \to \pm \infty$, 
% % % one would expect the fields to coincide. 

Test particles moving according to this field may have worldlines substantially different 
from those predicted by the Maxwell theory.
In the case of a source undergoing hyperbolic motion, 
the test particles can be shown to move highly above their mass shell.
                                                      
Generally, a particle may follow any motion in spacetime, including spacelike motion, 
and reverse timelike motion,
which would be interpreted by an observer as an anti-particle moving forward in time.
The off-shell motion can be shown to result from \emph{transfer of mass} to and from test particles by the field.
The reverse temporal motion of a particle can be viewed according to Stueckelberg's original view 
\cite{Stueckelberg1941,Stueckelberg1942} 
as \emph{a classical analog of the anti-particle}.

Since the wave equation is in 5D, 
its support is the \emph{entire past $\tau$ history} of the source\footnote{Applies to any odd dimensional spacetime.}.
The Huygens' principle does not apply, and thus, sources leave a trail, or a wake along their path.

Generally, the evaluation of the fields at any given 5D observation point $(x^{\mu},\tau)$
requires the entire history of the source inside the past cone with respect to this observation point.
As this rarely yields closed-form analytic solutions, numerical computation is an essential tool in 
this investigation.

This work continues a previous study \cite{horjig2006} on the fields generated by a point source in \emph{uniform motion}.

The paper is organized as follows:
\begin{enumerate}
 \item In section \ref{sec:fundamentals}, we give a short overview of the manifestly covariant relativistic dynamics of Stueckelberg 
       and the corresponding off-shell electrodynamics.
 \item In \ref{sec:retarded_green_functions} we provide the general $\tau$ retarded Green-Functions. These are derived
       from a \emph{ultrahyperbolic} generalized Riemann-Liouville integro-differential operator, 
       taken from Nozaki \cite{Nozaki_1964}.
 \item In \ref{sec:conditions_for_regularizability} we discuss conditions in which the fields are regularizable.
 \item In \ref{sec:numerics_and_results} we show the numerical results, and qualitatively compare the solution to the 
       Maxwell case. 
 \item In the last section \ref{sec:summary} we provide a summary and future prospects.
 \item We also provide an appendix for a more detailed derivation of the $\tau$-retarded Green function,
       and the associated regularization.
\end{enumerate}

\section{Fundamentals}
    \label{sec:fundamentals}
    %%
%% File   : UAP_Fundamentals.tex
%% Purpose: Overview of Stueckelberg relativistic dynamics + electrodynamics.
%%
%%
%%<===================MAXIMAL LINE LENGTH==========================>

% =============================================================== %
% --------------------------------------------------------------- %
% ----------------------> Fundamentals <------------------------- %
% --------------------------------------------------------------- %
% =============================================================== %
\subsection{Stueckelberg mainfestly covariant relativistic dynamics}
A so-called offshell classical and quantum electrodynamics has been
constructed \cite{SaadHorArsh1989} from a fundamental theory of
relativistic dynamics of 4D particles, termed \emph{events}, in a
framework first derived by Stueckelberg
\cite{Stueckelberg1941,HorPir1973}.

Stueckelberg defined a Lorentz invariant Hamiltonian-like generator
of evolution, over 8D phase space, with dynamics parameterized by a
Lorentz invariant $\tau$, in both classical and quantum relativistic
mechanics. Solutions of the relativistic quantum two body bound
state problem, defined in terms of an invariant potential function,
agree (up to relativistic corrections) with solutions of the
non-relativistic Schr\"odinger equation \cite{ArshHor1989_1,ArshHor1989_2,ArshHor1989_3}. The
experiments of Lindner, \emph{et. al.} \cite{Lindner2005}, moreover,
showing quantum interference in time can be explained in a simple
and consistent way in the framework of this theory
\cite{Horwitz2005}, and provides strong evidence that the time $t$,
should be regarded as a quantum observable, as required in this
framework.

In the classical manifestly covariant theory
\cite{Stueckelberg1941,HorPir1973}, the Hamiltonian of a free
particle is given by
\begin{align}
    \label{eq:stueckelberg_classical}
    K = \dfrac{p_{\mu} p^{\mu}}{2M}
\end{align}
where $x^{\mu} = [ct, \vb{x}]$ and $p^{\mu} = [E/c, \vb{p}]$. A
simple model for an interacting system is provided by the potential
model
\begin{align}
    \label{eq:stueckelberg_classical_with_V}
    K & =
        \dfrac{p_{\mu} p^{\mu}}{2M}
        +
        V(x)
\end{align}
The Hamilton equations are (the dot indicates derivative with respect to the independent parameter $\tau$)
\begin{align}
    \label{eq:hamilton_equations_classical_K}
    \dot{x}^{\mu} =  \dfrac{\partial K}{\partial p_{\mu}} =  \dfrac{1}{M} p^{\mu} \qquad \qquad
    \dot{p}^{\mu} = -\dfrac{\partial K}{\partial x_{\mu}} = -\dfrac{\partial V}{\partial x^{\mu}}
\end{align}
It follows from \eqref{eq:hamilton_equations_classical_K} that
\begin{align*}
    \vb{v}
    & =
        \dfrac{d\vb{x}}{dt}
    =
        \dfrac{\dot{\vb{x}}}{\dot{t}}
    =
        \dfrac{\vb{p}}{E}
\end{align*}
which is the standard formula obtained for velocity in special
relativity (we take $c=1$ in the following).

Horwitz and Piron \cite{HorPir1973} generalized the framework to
many-body systems, and gave $\tau$ the physical meaning of a
\emph{universal historical time}, correlating events in spacetime.

The general many-body, $\tau$ invariant, classical evolution
function is defined as
\begin{align}
    \label{eq:Stueckelberg_Classical_Hamiltonian_Many_Particles}
    K = \sum_{\TT n = 1}^{N}
        \dfrac{1}{2M_{n}} \eta_{\mu \nu} p_{n}^{\mu} p_{n}^{\nu} +
        V(x_{1}, x_{2}, ... , x_{N})
\end{align}
where $\eta_{\mu \nu} = diag(-,+,+,+)$ and $n$ sums over all
particles of the system, and, in this case, we have taken the
potential function $V$ not to be a function of momenta or $\tau$.
The classical equations of motion, for a single particle system in
an external potential $V(x)$, are similar to the non-relativistic
Hamilton equations, 
\begin{align}
    \dot{x}_{n}^{\mu} =  \dfrac{\partial K}{\partial p_{n \; \mu}} =   \dfrac{1}{M_n} p_{n}^{\mu}                    \qquad \qquad
    \dot{p}_{n \; \mu} = -\dfrac{\partial K}{\partial x_{n}^{\mu}} = - \dfrac{\partial V}{\partial x_{n}^{\mu}}
\end{align}
In the usual formulation of relativistic dynamics (cf.
\cite{Rindler1991}), the energy-momentum is constrained to a
\emph{mass-shell} defined as
\begin{align}
    p^{\mu} p_{\mu} = \mathbf{p}^2 - E^2 = - M^2
\end{align}
where $M$ is a given fixed quantity, a property of the particle. In
the Stueckelberg formulation, however, the event mass is generally
unconstrained. Since in \eqref{eq:stueckelberg_classical_with_V},
the value of $K$ is absolutely conserved, $p_{\mu} p^{\mu} = -m^2$
is constant only in the special case where
\begin{align*}
    \dfrac{d}{d\tau} V(x) =
        \left[
            \dot{x}^{\mu}
            \dfrac{\partial}{\partial x^{\mu}}
            +
            \dfrac{\partial}{\partial \tau}
        \right]
        V(x) =
        \dot{x} \cdot \mathbf{\nabla} V(x) = 0
\end{align*}
In this case, the particle remains in a specific mass shell, which
may or may not coincide with its so-called \emph{Galilean target
mass}, usually denoted by $M$\footnote{In the non-relativistic
limit, the mass distribution converges to a single point; one may
choose the parameter $M$ to have this \emph{Galilean target mass
value} \cite{Burakovsky1996}. We shall assume that $M$ has this
value in the following. }. In the general case, however, $p^{\mu}
p_{\mu} \equiv -m^2$ is a dynamical (Lorentz invariant) property,
which may depend on $\tau$. The relation between $\tau$ and the
proper time $s$, in the model of eq.
\eqref{eq:Stueckelberg_Classical_Hamiltonian_Many_Particles}, is
given by
\begin{align}
    ds^2 & \equiv - dx^{\mu} dx_{\mu}     = - \dot{x}^{\mu} \dot{x}_{\mu} d\tau^2 =
                  - \dfrac{1}{M^2} p^{\mu} p_{\mu} d\tau^2 =
                  \dfrac{m^2}{M^2} d\tau^2
\end{align}
Thus, the proper time $ds$, and \emph{universal time} $d\tau$, are
related through the ratio between the dynamical Lorentz invariant
mass $m$, and the \emph{Galilean target mass} $M$. If $V(x)$ goes to
zero asymptotically, then it becomes constant. Since this asymptotic
value is usually what is measured in experiment, we may assume that
it takes on the value of the Galilean target mass. Although there
are no detailed models at present, one assumes that there is a
stabilizing mechanism (for example, self-interaction or, 
in terms of a minimal free energy in 
statistical mechanics as, for example, in condensation phenomena \cite{Burakovsky1996}) 
which brings the particle, at least to a good
approximation, to a defined mass value, such that
\begin{align*}
    K = \dfrac{1}{2M} p^{\mu} p_{\mu} = \dfrac{-m^2}{2M} = - \dfrac{M}{2}
\end{align*}

For the quantum case, for which $p^{\mu}$ is represented by 
$-i\partial / \partial x_{\mu}$, the Stueckelberg Schr\"odinger equation
is taken to be (we take $\hbar = 1$ in the following)
\begin{align}
    \label{eq:stueckelberg_hamiltonain_quantum}
    i \dfrac{\partial \Psi_{\tau}(x)}{\partial \tau} = K \Psi_{\tau}(x)
\end{align}
The Stueckelberg classical and quantum relativistic dynamics have
been studied for various systems in some detail, including the
classical relativistic Kepler problem \cite{HorPir1973} and, as
mentioned above, the quantum two body problem for a central
potential \cite{ArshHor1989_2}.

% ================================================================================================================= %
% ----------------------------------------------------------------------------------------------------------------- %
% ----------------------> OSE Electrodynamics <-------------------------------------------------------------------- %
% ----------------------------------------------------------------------------------------------------------------- %
% ================================================================================================================= %
\subsection{Off-Shell Electrodynamics}
What we shall call "pre-Maxwell" off-shell electrodynamics is
constructed in a similar fashion to the construction of standard
Maxwell electrodynamics from the Schr\"odinger equation
\cite{SaadHorArsh1989}.

Under the local gauge transformation
\begin{align}
    \label{eq:gauge_transformation_definition}
    \Psi'_{\tau}(x) = e^{- i e_0 \chi(x,\tau)} \Psi_{\tau}(x)
\end{align}
5 compensation fields $a^{\alpha}(x,\tau)$ ($\alpha \in
\{0,1,2,3,5\}$) are implied, such that with the transformation
\begin{align*}
    a'_{\TT \alpha}(x,\tau) = a_{\alpha}(x,\tau) - \partial_{\alpha} \chi(x,\tau)
\end{align*}
the following modified Stueckelberg-Schr\"odinger equation remains form invariant
\begin{align}
    \label{eq:stueckelberg_schrodinger_after_gauge}
    \left[ i \dfrac{\partial}{\partial \tau} + e_0 a_{5}(x,\tau) \right] \Psi_{\tau}(x) =
    \dfrac{1}{2M}
    \left[
        (p^{\mu} - e_0 a^{\mu})
        (p_{\mu} - e_0 a_{\mu})
    \right]
    \Psi_{\tau}(x)
\end{align}
under the transformation \eqref{eq:gauge_transformation_definition}.

We can see this by observing the following relations:
\begin{align*}
    \left[ p_{\mu} - e_0 a'_{\mu} \right] \Psi' & =
        \left[- i \dfrac{\partial}{\partial x^{\mu}} - e_0 \left(a_{\mu} - \dfrac{\partial}{\partial x^{\mu}} \chi \right) \right]
        e^{-i e_0 \chi} \Psi = \\
    & =
        \left[
            - e_0 \dfrac{\partial}{\partial x^{\mu}} \chi  - i \dfrac{\partial \Psi}{x^{\mu}} -
            e_0 \left(a_{\mu} - \dfrac{\partial \chi}{\partial x^{\mu}} \right)
        \right]
        e^{-i e_0 \chi} \Psi  = \\
    & =
        e^{-i e_0 \chi}  [p_{\mu} - e_0 a_{\mu}] \Psi \\
        \\
    \intertext{and}
    \left[
        i \dfrac{\partial}{\partial \tau}
        +
        e_0 (a_{5} - \dfrac{\partial \chi}{\partial \tau})
    \right]
    e^{-i e_0 \chi} \Psi
    & =
        \left[
            e_0 \dfrac{\partial \chi}{\partial \tau}
            +   i \dfrac{\partial \Psi}{\partial \tau}
            +   e_0 ( a_{5} - \dfrac{\partial \chi}{\partial \tau} )
        \right]
            e^{- i e_0 \chi} \Psi = \\
    & =
        e^{-i e_0 \chi}
        \left[
            i \dfrac{\partial}{\partial \tau}
            +   e_0 a_5
        \right]
        \Psi
\end{align*}
The result is then, of the same form as for the usual $U(1)$ gauge
compensation argument for the non-relativistic Schr\"{o}dinger
equation. Thus, the classical (and quantum) evolution function for a
particle, under an external field, is then given by
\begin{align}
    K =
        \dfrac{1}{2M}
        \left[
            p - e_0 a(x,\tau)
        \right]^2
        -
        e_0 a^{5}(x,\tau)
\end{align}
(where we have used the shorthand notation of $x^2 = x_{\mu} x^{\mu}$)
and the corresponding Hamilton equations are
\begin{align}
    \label{eq:Hamilton_Equations_1}
    \dot{x}^{\mu}(\tau) & =
        \dfrac{\partial K}{\partial p_{\mu}} =
        \dfrac{1}{M}
        \left[
            p^{\mu} - e_0 a^{\mu}
        \right] \\
    \label{eq:Hamilton_Equations_2}
    \dot{p}^{\mu}(\tau) & =
        - \dfrac{\partial K}{\partial x_{\mu}} =
        \dfrac{e_0}{M}
        \left(
            p - e_0 a(x,\tau)
        \right)_{\nu}
        \partial^{\mu} a^{\nu}(x,\tau) +
        e_0 \partial^{\mu} a^{5}(x,\tau)
\end{align}
Here, $e_0$ is proportional to the Maxwell charge $e$ through a
dimensional constant, which is discussed below. Second order
equations of motion for $x^{\mu}(\tau)$, a generalization of the
usual Lorentz force, follow from the Hamilton equations
\eqref{eq:Hamilton_Equations_1} and \eqref{eq:Hamilton_Equations_2}
\cite{SaadHorArsh1989}
\begin{align}
    \label{eq:5D_Lorentz_force}
    M \ddot{x}^{\mu} = e_0 \dot{x}^{\nu} f_{\; \nu}^{\mu} + e_0 f_{\; 5}^{\mu}
\end{align}
where for $\alpha,\beta = 0,1,2,3,5$ the antisymmetric tensor
\begin{align}
    \label{eq:ose_fields_from_potentials}
    f^{\alpha \beta} \equiv \partial^{\alpha} a^{\beta} - \partial^{\beta} a^{\alpha}
\end{align}
is the (gauge invariant) 5D field tensor. Moreover, second order
wave equation for the fields $f^{\alpha \beta}$ can be derived from
a Lagrangian density as follows \cite{SaadHorArsh1989}:
\begin{align}
    \label{eq:ose_field_langarangian}
    \mathscr{L} =  -\dfrac{\lambda}{4} f_{\alpha \beta} f^{\alpha \beta} - e_0 a_{\alpha} j^{\alpha}
\end{align}
which produces the wave equation
\begin{align}
    \label{eq:ose_fields_wave_equation}
    \lambda \partial_{\alpha} f^{\beta \alpha} = e_0 j^{\beta}
\end{align}
$\lambda$ is a dimensional constant, which will be shown below to have dimensions of length.
The sources $j^{\beta}(x,\tau)$ depend both on spacetime and on $\tau$, and obey
the continuity equation
\begin{align}
    \label{eq:continuity}
    \partial_{\alpha} j^{\alpha} = \partial_{\mu} j^{\mu} + \partial_{\tau} \rho = 0
\end{align}
where $j^{5} \equiv \rho$ is a Lorentz invariant \emph{spacetime
density of events}. This equation follows from
\eqref{eq:stueckelberg_schrodinger_after_gauge} for
\begin{align*}
    \rho_{\tau}(x) & = \Psi^{*}_{\tau}(x) \Psi_{\tau}(x)
    \\
    j^{\mu}_{\tau}(x) & =
        - \dfrac{i}{2M}
        \left[
            \Psi^{*}_{\tau}(x)
            \left(
                i \partial^{\mu} - e_{0} a^{\mu}(x,\tau)
            \right)
            \Psi_{\tau}(x)
            +
            c.c.
        \right]
\end{align*}
as we discuss below, and also the classical from the argument given
below.

\subsubsection{Currents of point events}
Jackson \cite{Jackson1995} showed that a conserved current for a
moving point charge can be derived in a covariant way by defining
the current as
\begin{align}
    \label{eq:maxwell_current_of_point_particle}
    J^{\mu}(x) = e \int_{-\infty}^{+\infty} \ud s \, \dot{z}^{\mu}(s) \delta^4 [x - z(s)]
\end{align}
In this case, $s$ is the proper time, and $z^{\mu}(s)$ the
world-line of the point charge (for free motion, $s$ may coincide
with $\tau$), and $\dot{z}^{\mu}(s) = \dfrac{d}{ds} z^{\mu}(s)$.
Then,
\begin{align}
    \label{eq:maxwell_current_conservation_proof}
    \partial_{\mu} J^{\mu} =
        - e \int_{-\infty}^{+\infty} \ud s \, \dfrac{\ud }{\ud s} \delta^4 [x - z(s)] =
        - e \, \lim \limits_{L \rightarrow +\infty} \delta^4 [x - z(s)] \Bigg|_{-L}^{+L}
\end{align}
which vanishes if $z^{\mu}(s)$ (or, for example, just the time
component $z^{0}(s)$) becomes infinite for $s \rightarrow \pm
\infty$, and the observation point $x^{\mu}$ is restricted to a
bounded region of spacetime, e.g., the laboratory. We therefore,
with Jackson, identify $J^{\mu}$ as the Maxwell current. We see that
this current is a \emph{functional} on the world line, and the usual
notion 
of a "particle" associated with a conserved 4-current (and 
therefore a charge charge corresponding to the space integral of the 
fourth component), 
corresponds to this functional on the world line.

If we identify $\delta^4 [x - z(s)]$ with a density $\rho_{s}(x)$
and the local (in $\tau$) current $\dot{z}^{\mu}(s) \delta^4 [x -
z(s)]$ with a local current $j^{\mu}_{s}(x)$
\begin{align}
    \label{eq:ose_current_of_point_event}
    \rho_{s}(x)      = \delta^4 [x - z(s)] \qquad \qquad
    j^{\mu}(x,s) = \dot{z}^{\mu}(s) \delta^4 [x - z(s)]
\end{align}
then the relation
\begin{align*}
    \dfrac{d}{ds} \delta^4 [x - z(s)] = -\dot{z}^{\mu}(s) \partial_{\mu} \delta^4 [x - z(s)]
\end{align*}
used in the above demonstration in fact corresponds to the
conservation law (reverting to the more general parameter $\tau$ in
place of the proper time $s$) \eqref{eq:continuity}
\begin{align}
    \label{eq:ose_current_conservation}
    \partial_{\mu} j^{\mu}(x,\tau) + \partial_{\tau} \rho(x,\tau) = 0
\end{align}
What we call the \emph{pre-Maxwell} current of a point \emph{event}
is then defined as
\begin{align}
    \label{eq:point_source_current}
    j^{\alpha}(x,\tau) = \dot{z}^{\alpha}(\tau) \delta^4 [x - z(\tau)]
\end{align}
where $j^{5}(x,\tau) \equiv \rho(x,\tau)$ and $\dot{z}^{5}(\tau)
\equiv 1$ (since $z^{5}(\tau) \equiv \tau$). Integrating
\eqref{eq:ose_fields_wave_equation} over $\tau$, we recover the
standard Maxwell equations for Maxwell fields defined by
\begin{align}
    A^{\mu}(x) & = \int a^{\mu}(x,\tau) \ud\tau
    \label{eq:A_from_a}
\end{align}
We therefore call the fields $a^{\mu}(x,\tau)$ \emph{pre-Maxwell
fields}. From \eqref{eq:ose_field_langarangian}
\begin{align}
    \lambda
    \int_{-\infty}^{+\infty}
        \left[
            \partial_{\mu} f^{\mu \nu}(x,\tau)
            +
            \partial_{5} f^{\mu 5}(x,\tau)
        \right]
        \ud\tau
    & =
        \lambda
        \int_{-\infty}^{+\infty}
            \partial_{\alpha} f^{\mu \alpha}(x,\tau)
        \ud\tau
    \nonumber
    \\
    & =
        e_{0} \int_{-\infty}^{+\infty} j_{\beta}(x,\tau) \, \ud\tau
    \nonumber
    \\
    & =
        e_{0} \dfrac{1}{e} J^{\mu}(x)
    \label{eq:derivation_of_A_from_a}
\end{align}
where the $\tau$ integral of $\partial_{5} f^{\mu 5}$ vanishes for
local sources. Therefore, \eqref{eq:derivation_of_A_from_a} proves
\eqref{eq:A_from_a}. Moreover,
\begin{align}
    \partial_{\nu} F^{\mu \nu}
    & =
        \dfrac{e_{0}}{\lambda e}
        J^{\mu}(x)
\end{align}
we find the constants $\lambda$ and $e_0$ are related by
\begin{align}
    \lambda & = \dfrac{e_{0}}{e}
    \\
    e       & = \dfrac{e_{0}}{\lambda}
\end{align}
where $e$ is the standard Maxwell charge.

For the quantum theory, a real positive definite density function $\rho_{\tau}(x)$
can be derived from the Stueckelberg-Schr\"odinger equation \eqref{eq:stueckelberg_hamiltonain_quantum}
\begin{align}
    \rho_{\tau}(x) = |\Psi_{\tau}|^2 = \Psi^{*}_{\tau} (x) \Psi_{\tau}(x)
\end{align}
which can be identified with the $\rho(x,\tau) = \delta^4[x -
z(\tau)]$ in the classical (relativistic) limit. It follows from the
Stueckelberg equation
\eqref{eq:stueckelberg_schrodinger_after_gauge} that the continuity
equation \eqref{eq:ose_current_conservation} is then satisfied for
the gauge invariant current
\begin{align}
    j^{\mu}_{\tau}(x) =
        - \dfrac{1}{2M}
        \left[
            \Psi^{\ast}_{\tau}(x) (i \partial^{\mu} - e_0 a^{\mu}(x,\tau)) \Psi_{\tau}(x)  +
            \text{c.c.}
        \right]
\end{align}
From \eqref{eq:ose_field_langarangian}, also valid for the quantum
theory \cite{SaadHorArsh1989,LandShnerbHorwitz1995}, and
\eqref{eq:A_from_a}, we find
\begin{align}
    \label{eq:ose_maxwell_current_integral}
    J^{\mu}(x) = e \int_{-\infty}^{+\infty} j^{\mu}(x,\tau) \, \ud\tau
\end{align}

\subsubsection{The wave equation}
From equations \eqref{eq:ose_fields_wave_equation} and \eqref{eq:ose_fields_from_potentials} one can derive
the wave equation for the potentials $a^{\TT\alpha}(x,\tau)$:
\begin{align}
    \label{eq:ose_potentials_wave_equation_before_gauge}
    \lambda \partial_{\beta} \partial^{\beta} a^{\alpha} -
    \lambda \partial^{\alpha} (\partial_{\beta} a^{\beta}) = e_0 \, j^{\alpha}
\end{align}
Under the generalized Lorentz gauge $\partial_{\beta} a^{\TT\beta} =
0$, the wave equation takes the simpler form
\begin{align}
    \label{eq:ose_potentials_wave_equation_after_gauge}
    \lambda \partial_{\beta} \partial^{\beta} a^{\alpha} =
    \lambda \left[ \Box^2 a^{\alpha} + \sigma_5 \dfrac{\partial^2 a^{\alpha}}{\partial \tau^2} \right] =
    e_0 \, j^{\alpha}(x,\tau)
\end{align}
where a $5^{th}$ diagonal metric component can take either signs
$\sigma_5 = \pm 1$, corresponding to $O(4,1)$ and $O(3,2)$
symmetries of the homogeneous field equations, respectively.

Integrating \eqref{eq:ose_potentials_wave_equation_after_gauge} with respect to $\tau$,
and assuming that $\lim\limits_{\tau \rightarrow \pm \infty} \partial_{\tau} a^{\alpha}(x,\tau) = 0$,
we obtain
\begin{align*}
    \lambda
    \int_{-\infty}^{+\infty} \ud\tau \,
    \left[ \Box^2 a^{\alpha} + \sigma_5 \dfrac{\partial^2 a^{\alpha}}{\partial \tau^2} \right] =
    \dfrac{e_0}{e} J^{\alpha}(x)
\end{align*}
Identifying
\begin{align}
    \label{eq:maxwell_from_pre_maxwell_potential}
    A^{\mu}(x) = \int_{-\infty}^{+\infty} \ud\tau \, a^{\mu}(x,\tau)
\end{align}
we obtain
\begin{align*}
    \lambda \Box^2 A^{\mu}(x) & = \dfrac{e_0}{e} J^{\mu}(x)
    \\
    \intertext{i.e.}
    \Box^2 A^{\mu}(x) & = J^{\mu}(x)
\end{align*}
(for $\mu = 0,1,2,3$)

Therefore, the Maxwell electrodynamics is properly contained in the
5D electromagnetism.

\subsubsection{A note about units}
In natural units ($\hbar = c = 1$), the Maxwell potentials $A^{\mu}$ have units of $1/L$.
Therefore, the pre-Maxwell OSE potentials $a^{\alpha}$ have units of $1/L^2$, and
in order to maintain the action integral
\begin{align}
    S = \int_{-\infty}^{+\infty} \mathscr{L} \, \ud\tau \, \ud^4x
\end{align}
dimensionless, the coefficient $\lambda$ in \eqref{eq:ose_field_langarangian} must have units of $L$, forcing
$e_0$ to have units of $L$ as well  (hence $e$ is dimensionless).

The Fourier transform of the pre-Maxwell OSE fields
\begin{align}
    \tilde{a}^{\mu}(x,s) = \int_{-\infty}^{+\infty} e^{is\tau} a^{\mu}(x,\tau) \, \ud\tau
\end{align}
and equation \eqref{eq:maxwell_from_pre_maxwell_potential} suggest that the Maxwell potentials and fields
correspond to the \emph{zero mode} of the pre-maxwell OSE fields, with respect to $\tau$, i.e.,
\begin{align}
    \label{eq:maxwell_field_from_zero_mode}
    A^{\mu}(x) = \tilde{a}^{\mu}(x,s) |_{s = 0}
\end{align}

% ================================================================================================================= %
% ----------------------------------------------------------------------------------------------------------------- %
% ----------------------> Solutions of the equations <------------------------------------------------------------- %
% ----------------------------------------------------------------------------------------------------------------- %
% ================================================================================================================= %
\subsection{Solutions of the wave equation}
In \cite{horjig2006}, we have shown the Green-functions (GF)
associated with the 5D wave-equation \eqref{eq:ose_potentials_wave_equation_after_gauge}  
that are consistent with uniformly moving point sources are 
\begin{align}
    \label{eq:2006_green_function}
    g_{\sigma_{5}}(x,\tau) =
        \dfrac{\sigma_{5}}{4\pi^2}
        \lim\limits_{\epsilon \to 0^{+}}
        \dfrac{\partial}{\partial \epsilon}
            \dfrac{\theta[ - \sigma_{5} \left(x^2 + \sigma_{5} \tau^2 \right) + \epsilon ] }
                  {\sqrt { - \sigma_{5} \left(x^2 + \sigma_{5} \tau^2 \right) + \epsilon } }
\end{align}
where $\sigma_{5} = \pm 1$ for the sign of $\tau$ in the metric, $\pm 1$ for $(4,1)$ and $(3,2)$ respectively,
obeying the wave equation on for a $\delta^{5}(x,\tau)$ source
\begin{align}
    \label{eq:green_function_definition}
    \partial_{\beta} \partial^{\beta} a^{\alpha}(x,\tau)
    & = 
        \delta^4(x) \delta(\tau)
\end{align}

Using the GF's \eqref{eq:2006_green_function} 
the general field generated by a given source can then be found by
integration on its support
\begin{align}
    \label{eq:hps_potential_solution_using_green_function}
    a^{\alpha}(x,\tau) =
            e_0
            \int \ud^4x' \, \ud\tau'
                \,
                g(x-x',\tau-\tau') \, j^{\alpha}(x',\tau')
\end{align}
and applying it to a point particle given by \eqref{eq:ose_current_of_point_event}.
The potentials of point events are then given as
\begin{align}
    \label{eq:hps_potential_solution_using_green_function_for_point_particle}
    a^{\alpha}(x,\tau) =
            e_0
            \int_{-\infty}^{+\infty}
                \ud\tau'
                \,
                g(x-z(\tau'),\tau-\tau') \, \dot{z}^{\alpha}(\tau')
\end{align}

However, in the framework of off-shell electrodynamics, the fields are to be causal in $\tau$,
and since \eqref{eq:2006_green_function} is \emph{symmetric} with respect to $\tau$, they
cannot be used for non-inertial (hence, non-uniform) source motion, as the notion of flow
of radiation in $\tau$ is not properly accounted for (cf. \cite{Poisson:1999tv}).
Therefore, the next section is focused on achieving consistent $\tau$-retarded GF's.

\section{Retarded Green Functions}
    \label{sec:retarded_green_functions}
    %%
%% File   : UAP_green_functions.tex
%% Purpose: Give *short* derivation of the tau retarded green-functions.
%%          (the long derivation is in the appendix)
%%
\subsection{Retarded ultrahyperbolic Green-Functions}
The generalization of \emph{potential theory} to a spacetime with $(p,q)$ signature was achieved
by Nozaki \cite{Nozaki_1964}, that extended earlier work of Riesz \cite{riesz_1938,riesz_1949}, on its own a generalization
of potential theory to the Lorentzian space $(p,1)$.

It begins with a definition of a ultrahyperbolic distance between two points $x,y \in \mathbb{R}^{p,q}$ 
\begin{align}
    \label{eq:p_q_distance}
    r^2(x-y)
    & \equiv
        \eta_{ij}^{(p,q)}(x^{i} - y^{i})(x^{j} - y^{j})
\end{align}
where $\eta^{(p,q)}_{ij}$ is the corresponding $\mathbb{R}^{p,q}$ metric 
\begin{align}
    \label{eq:p_q_metric_tensor}
    \eta^{(p,q)}_{ij}
    & = 
        \text{diag}
        (
            \overbrace{+1, \ldots, +1}^{p} , 
            \overbrace{-1, \ldots, -1}^{q} 
        )
\end{align}

Then, the ultrahyperbolic inverted hypercone $\mathcal{D}_{x}$ with a vertex 
at $x \in \mathbb{R}^{p,q}$ is given by
\begin{align}
    \label{eq:normal_retarded_hypercone_definition}
    \mathcal{D}_{x} & = 
        \left\{
                \xi \in \mathbb{R}^{p,q}
            \Big|
                \quad
                r^{2}(x - \xi) 
                = 
                \eta^{(p,q)}_{ij} (x^{i} - \xi^{i}) (x^{j} - \xi^{j}) > 0
                ,
                \quad
                x_{1} - \xi_{1} > 0
        \right\}
\end{align}
$\mathcal{D}_{x}$ is in fact, the retarded hyper-cone, with respect to the first coordinate $x_{1}$.

A generalized \emph{Riemann-Liouville} integration operator $J^{\alpha}$ is then defined by
\begin{align}
    \label{eq:nozaki_J_alpha}
    J^{\alpha} f(x)
    & = 
        \int_{\mathcal{D}_{x}}
            \dfrac{r^{\alpha - m}(y) }
                  {K_{m}(\alpha)     }
            f(x-y)
            \,
            \ud y
    =
        \Phi_{\alpha}(y) * f(x-y)
\end{align}
where $m = p+q$ is the dimensionality of the entire ultrahyperbolic space, and
\begin{align}
    r^{a}(x) 
    & = 
        \left[r^{2}(x)\right]^{a/2} 
\end{align}
and $*$ denotes \emph{convolution}, where
\begin{align}
    \label{eq:definition_phi}
    \Phi_{\alpha}(y)
    & \equiv
        \dfrac{r^{\alpha - m}(y)}
              {K_{m}(\alpha)    }
\end{align}

$K_{m}(\alpha)$ is a normalization constant that is determined by fixing $J^{\alpha}$ such that 
the function $f(x) = e^{x_{1}}$ remains stationary under the action of $J^{\alpha}$.

Furthermore, 
\begin{align}
    \label{eq:normalization}
    K_{m}(\alpha)
    & = 
        \pi^{(m-1)/2}
        \dfrac
                %% numerator %%
                {
                    \Gamma
                        \left(
                            \frac{\alpha - m + 2}
                                 {       2      }
                        \right)
                    \Gamma
                        \left(
                            \alpha
                        \right)
                    \Gamma
                        \left(
                            \frac{1 - \alpha    }
                                 {    2         }
                        \right)
                }
                %% denominator %%
                {
                    \Gamma
                        \left(
                            \frac{\alpha - p + 2}
                                 {       2      }
                        \right)
                    \Gamma
                        \left(
                            \frac{p - \alpha    }
                                 {    2         }
                        \right)
                }
\end{align}
$K_{m}(\alpha)$ can be thought as a \emph{ultrahyperbolic generalization of the $\Gamma$-function}.
Using its analytic properties, it can be shown that \cite{Nozaki_1964}
\begin{align*}
    \lim\limits_{\alpha \to 0^{+}} J^{\alpha} f(x) 
    & = 
        f(x)
\end{align*}
and thus
\begin{align}
    \label{eq:phi_alpha_limit_to_delta_function}
    \lim\limits_{\alpha \to 0^{+}} \Phi_{\alpha}(x) & = \delta(x)
\end{align}

Furthermore, the ultrahyperbolic wave-operator given by
\begin{align}
    \label{eq:ultrahyperbolic_wave_operator}
    \Delta 
    & \equiv
        \eta^{ij}_{(p,q)} \partial_{i} \partial_{j}
\end{align}
can be shown to obey
\begin{align*}
    \Delta J^{\alpha} f(x) 
    & = 
        J^{\alpha-2} f(x)
\end{align*}
Or
\begin{align}
    \label{eq:wave_operator_on_phi_alpha}
    \Delta \Phi_{\alpha}(x) & = \Phi_{\alpha-2}(x)
\end{align}

Using \eqref{eq:wave_operator_on_phi_alpha} and \eqref{eq:phi_alpha_limit_to_delta_function}, 
it can easily be seen that the \emph{fundamental-solution} for the wave operator is 
\begin{align*}
    \Delta \lim\limits_{\alpha \to 0^{+}} \Phi_{\alpha+2}(x) & = \Phi_{0}(x)
\end{align*}
Thus
\begin{align}
    \label{eq:nozaki_fundamental_solution}
    g_{(p,q)}(x) & = \Phi_{2}(x)
                   = \dfrac{r_{+}^{2-m}(x)}
                           {K_{m}(2)      }
\end{align}

\subsection{Modified retarded ultrahyperbolic Green-Functions}
Equation \eqref{eq:2006_green_function} is based a simple Principal Part solution based on Fourier analysis \cite{horjig2006} 
of \eqref{eq:green_function_definition},
and carries a negative signature for $\tau^2$ term in both $(4,1)$ and $(3,2)$ metrics.
We therefore seek a general retarded ultrahyperbolic fundamental solution 
on a hyper-cone for which retardation vertex is determined
by $x_{m}$\footnote{Recall that $m = p+q$ is the dimensionality of the overall system.}.
%%%which can be accomplished by analytic continuation of \eqref{eq:nozaki_fundamental_solution}.
In the appendix \ref{sec:derivation_of_retarded_green_function}, we apply a procedure
similar to Nozaki's to obtain the required opposite signature retardation
(this result may also be obtained by analytic continuation of \ref{sec:derivation_of_retarded_green_function}).

Beginning with the \emph{definition} of the retarded inverted hypercone in $(p,q)$ with \emph{vertex} at $x \in R^{p,q}$ by:
\begin{align}
    \label{eq:jigal_retarded_hypercone_definition}
    \mathcal{D'}_{x} & = 
        \left\{
                \xi \in \mathbb{R}^{p,q}
            \Big|
                \quad
                r^{2}(x - \xi) 
                = 
                \eta^{(p,q)}_{ij} (x^{i} - \xi^{i}) (x^{j} - \xi^{j}) > 0
                ,
                \quad
                x_{m} - \xi_{m} > 0
        \right\}
\end{align}

We similarly define a modified ultrahyperbolic Riemann-Liouville operator $L^{\alpha}$ such that
\begin{align}
    \label{eq:jigal_ultrahyperbolic_riemann_liouville}
    L^{\alpha} f(x)
    & = 
        (f * \phi_{\alpha})(x)
    =
        \int_{\mathcal{D'}_{x}}
            \dfrac{r_{+}^{\alpha - m}(x-y)}{N_{m}(\alpha)}
            f(y)
            \,
            \ud y
\end{align}

where, similarly to Nozaki's construction of $K_{m}(\alpha)$,
\begin{align}
    \label{eq:jigal_normalization}
    N_{m}(\alpha)
    & = 
        2
        \pi^{(m-3)/2}
        \sin
            \left(
                \dfrac{\pi p}{2}
            \right)
        \Gamma
            \left(
                \dfrac{2 + \alpha - m}{2}
            \right)
        \Gamma
            \left(
                \dfrac{1 - \alpha}{2}
            \right)
        \Gamma
            \left(
                \alpha
            \right)
\end{align}
where in this case
\begin{align*}
    \Delta \Phi_{\alpha}(x)
    & = 
        \Delta 
            \dfrac{r_{+}^{\alpha - m}}{N_{m}(\alpha)}
    = 
        -
        \Phi_{\alpha - 2}(x)
\end{align*}
and thus, the desired retarded GF's are ($x \in \mathbb{R}^{p,q}$):
\begin{align}
    \label{eq:jigal_ultrahyperbolic_green_functions}
    g_{(p,q)}(x)
    & = 
        - \lim_{\alpha \to 2} \Phi_{\alpha}(x)
    = 
        \dfrac{r_{+}^{2 - m}}
              {2 \pi^{m/2 - 1} \sin\left(  \dfrac{\pi p}{2} \right) \Gamma \left( \dfrac{4 - m}{2} \right)}
\end{align}

Applying \eqref{eq:jigal_ultrahyperbolic_green_functions} to the case of $(4,1)$ or $(3,2)$ metrics, we find:
\begin{align}
    \label{eq:jigal_green_functions_for_4_1_and_3_2}
    g(x)
    & = 
        \dfrac{\theta(\tau)}{4 \pi^2}
        \begin{cases}
            (-1)
            \dfrac{\theta(-x^2 - \tau^2)      }
                  {      (-x^2 - \tau^2)^{3/2}}
            \qquad
            \qquad
            &
            (4,1)
        \\
            \dfrac{\theta(x^2 - \tau^2)      }
                  {      (x^2 - \tau^2)^{3/2}}
            &
            (3,2)
        \end{cases}
\end{align}
which complies with the GF's obtained in \cite{horjig2006}, aside from a factor of $2$, due to the specific
choice of \emph{retarded GF}. This is in exact analogy to the Maxwell GF's in which 
\begin{align*}
    G_{\text{P}}(x) & = \dfrac{1}{2} 
                               \left[
                                    G_{\text{ret}}(x) + G_{\text{adv}}(x)
                               \right]
\end{align*}

Following the convention \cite{LandHor1991,horjig2006}, we denote the sign of $\tau$ in the metric by $\sigma_{5}$, 
and define $R(x) = - \sigma_{5} ( x^2 + \sigma_{5} \tau^2)$, we then find
\begin{align*}
    g(x) 
    & = 
        - \sigma_{5}
        \dfrac{\theta(\tau)}{4 \pi^2}
        \dfrac{\theta(R(x))}{R^{3/2}(x)}
\end{align*}

When a source $j^{\alpha}(x)$ is given, we find the field
\begin{align}
    \label{eq:field_due_to_green_function}
    a^{\alpha}(x)
    & = 
        \int_{\mathcal{D}_{x,x'}}  g(x-x') j^{\alpha}(x') \ud x'
    =
        - \dfrac{\sigma_{5}}{4\pi^2}
        \int_{\mathbb{R}^{5}}  \theta(\tau - \tau') 
                               \dfrac{\theta(R(x - x'))}
                                     {R^{3/2}(x-x')    } 
                               \,
                               j^{\alpha}(x') \, dx'
\end{align}

The integration over regions where $R(x-x') = 0$ is defined in the sense of \emph{generalized-functions},
as given in \cite{Gelfand1964_1}, by employing \emph{canonical regularization}
(see appendix \ref{sec:canonical_regularization}).

For a point particle, $j^{\alpha}$ is 
\begin{align}
    \label{def:point_particle_current}
    j^{\alpha}(x,\tau) & = q \dfrac{dz^{\alpha}}{d\tau}  \delta^{4} \left( x - z(\tau) \right)
\end{align}
where $z^{\alpha}(\tau)$ is the spacetime position of the source charge, and by definition, $z^{5}(\tau) \equiv \tau$.
Inserting \eqref{def:point_particle_current} to \eqref{eq:field_due_to_green_function}, we find
\begin{align}
    \label{eq:field_due_to_point_source}
    a^{\alpha}(x)
    =
        - \dfrac{q \sigma_{5}}{4\pi^2}
        \mathcal{R}
        \int_{-\infty}^{\tau}  \dfrac{\theta\left( R(x - z(\tau'))   \right) }
                                     {R^{3/2}(x-z(\tau'))                    } 
                               \,
                               \dot{z}^{\alpha}(\tau')
                               \,
                               \ud\tau'
\end{align}
where we explicitly denoted regularization by $\mathcal{R}$.

Similarly, the fields can be found by taking the spacetime-$\tau$ derivatives
\begin{align}
    \label{eq:f_field_due_to_point_source}
    f^{\alpha \beta}(x)
    & =
        \partial^{\alpha} a^{\beta} - \partial^{\beta} a^{\alpha}
    =
        \dfrac{3 q \sigma_{5}}{8\pi^2}
        \mathcal{R}
        \int_{-\infty}^{\tau}  \dfrac{\theta\left( R \right) }
                                     {R^{5/2}                } 
                               \,
                               \left[
                                    \dot{z}^{\beta } \partial^{\alpha} R 
                                    -
                                    \dot{z}^{\alpha} \partial^{\beta } R 
                               \right]
                               \,
                               \ud\tau'
\end{align}
where $R$ and $z^{\alpha}$ are evaluated at $\tau'$.
Hence forward, we shall denote by $R(\tau')$ the expression
\begin{align}
    \label{eq:R_tau_prime}
    R(\tau')
    & = 
        - 
        \sigma_{5}
        (x_{\alpha} - z_{\alpha}(\tau'))
        (x^{\alpha} - z^{\alpha}(\tau'))
    =
        - 
        \sigma_{5}
        \left[
            (x - z(\tau'))^2 
            + 
            \sigma_{5}
            (\tau - \tau')^2
        \right]
\end{align}

\subsection{Regularization}
\label{subsec:regularization}
Integrals \eqref{eq:f_field_due_to_point_source} and \eqref{eq:field_due_to_point_source} are divergent 
at $\tau' = \tau'_{i}$, where $\tau'_{i}$ are the \emph{roots} of the equation $R(\tau'_{i}) = 0$
with the additional constraint of retardation $\tau'_{i} \leq \tau$. The divergence can be eliminated 
by following a procedure of \emph{canonical regularization} \cite{Gelfand1964_1} 
(see also appendix \ref{sec:canonical_regularization}). 
For example, let us assume that for a particular 5D observation $(x,\tau)$, the equation $R(\tau') = 0$
has a \emph{single root} $\tau'_{0}$ such that $R(\tau'_{0}) = 0$, and 
$R(\tau') < 0$ for $\tau' < \tau'_{0}$. 

Regularization of \eqref{eq:f_field_due_to_point_source}  can then be expressed as a sum of 3 operators, 
which are parameterized by a (small) $h$, as follows:
\begin{align}
    \label{def:canonical_regularization_for_case_1}
    \mathcal{R}
    & 
    \int_{\tau'_{0}}^{\tau}
            \dfrac{\phi^{\alpha \beta}(\tau')}
                  {R^{5/2}(\tau')            }
                   \,
                  \ud\tau'  
    = 
        \left[
            \mathcal{R}_{1}[h]
            +
            \mathcal{R}_{2}[h]
            +
            \mathcal{R}_{3}[h]
        \right]
            \int_{\tau'_{0}}^{\tau}
                    \dfrac{\phi^{\alpha \beta}(\tau')}
                          {R^{5/2}(\tau')            }
                           \,
                          \ud\tau'  
    \nonumber
    \\
    & = 
        \int_{\tau'_{0}}^{\tau'_{h}}
            \dfrac{\dot{R}(\tau') }
                  {R^{5/2}(\tau') }
            \left[
                \dfrac{\phi^{\alpha \beta}(\tau')}
                      {\dot{R}(\tau')            }
                -
                \dfrac{\phi^{\alpha \beta}(\tau'_{0})}
                      {\dot{R}(\tau'_{0})            }
                -
                R(\tau')
                \left(
                    \dfrac{\dot{\phi}^{\alpha \beta}(\tau'_{0})}
                          {\dot{R}^2(\tau'_{0})                }
                    -
                    \dfrac{\phi^{\alpha \beta}(\tau'_{0}) \ddot{R}(\tau'_{0}) }
                          {\dot{R}^3(\tau'_{0})                               }
                \right)
            \right]
            \ud\tau'
        +
    \nonumber
    \\ &
        \qquad 
        +
        \left[
            \dfrac{\phi^{\alpha \beta}(\tau'_{0}) }
                  {\dot{R}(\tau'_{0}) R_{h}^{3/2} }
            \left(
                1 
                -
                3 \dfrac{R_{h} \ddot{R}(\tau'_{0})}
                        {\dot{R}^2(\tau'_{0})     }
            \right)
            +
            3 \dfrac{\dot{\phi}^{\alpha \beta}(\tau'_{0})}
                    {\dot{R}^2(\tau'_{0}) R_{h}^{1/2}    }
        \right]
        +
    \nonumber
    \\ & 
        \qquad 
        +
        \int_{\tau'_{h}}^{\tau}
            \dfrac{\phi^{\alpha \beta}(\tau')}
                  {R^{5/2}(\tau')            }
                   \,
                  \ud\tau'  
\end{align}
where
\begin{align*}
%%    \label{def:phi_alpha_beta}
    \phi^{\alpha \beta}(\tau')
    & \equiv 
        \dot{z}^{\beta }(\tau') \partial^{\alpha} R(\tau') 
        -
        \dot{z}^{\alpha}(\tau') \partial^{\beta } R(\tau') 
    \\
    \tau'_{h} & = \tau'_{0} + h > \tau'_{0}
    \\
    R_{h} & = R(\tau'_{0} + h) 
    \\
    \dot{R}(\tau')
    & = 
        \dfrac{d}{d\tau'} R(\tau')
    \\
    \mathcal{R}_{1}[h]
    & 
    \int_{\tau'_{0}}^{\tau}
            \dfrac{\phi^{\alpha \beta}(\tau')}
                  {R^{5/2}(\tau')            }
                   \,
                  \ud\tau'  
        = 
        \int_{\tau'_{0}}^{\tau'_{h}}
            \dfrac{\dot{R}(\tau') }
                  {R^{5/2}(\tau') }
            \left[
                \dfrac{\phi^{\alpha \beta}(\tau')}
                      {\dot{R}(\tau')            }
                -
                \dfrac{\phi^{\alpha \beta}(\tau'_{0})}
                      {\dot{R}(\tau'_{0})            }
                -
                R(\tau')
                \left(
                    \dfrac{\dot{\phi}^{\alpha \beta}(\tau'_{0})}
                          {\dot{R}^2(\tau'_{0})                }
                    -
                    \dfrac{\phi^{\alpha \beta}(\tau'_{0}) \ddot{R}(\tau'_{0}) }
                          {\dot{R}^3(\tau'_{0})                               }
                \right)
            \right]
            \ud\tau'
    \\
    \mathcal{R}_{2}[h]
    & 
    \int_{\tau'_{0}}^{\tau}
            \dfrac{\phi^{\alpha \beta}(\tau')}
                  {R^{5/2}(\tau')            }
                   \,
                  \ud\tau'  
        = 
            \dfrac{\phi^{\alpha \beta}(\tau'_{0}) }
                  {\dot{R}(\tau'_{0}) R_{h}^{3/2} }
            \left(
                1 
                -
                3 \dfrac{R_{h} \ddot{R}(\tau'_{0})}
                        {\dot{R}^2(\tau'_{0})     }
            \right)
            +
            3 \dfrac{\dot{\phi}^{\alpha \beta}(\tau'_{0})}
                    {\dot{R}^2(\tau'_{0}) R_{h}^{1/2}    }
    \\
    \mathcal{R}_{3}[h]
    & 
    \int_{\tau'_{0}}^{\tau}
            \dfrac{\phi^{\alpha \beta}(\tau')}
                  {R^{5/2}(\tau')            }
                   \,
                  \ud\tau'  
        = 
        \int_{\tau'_{h}}^{\tau}
            \dfrac{\phi^{\alpha \beta}(\tau')}
                  {R^{5/2}(\tau')            }
                   \,
                  \ud\tau'  
\end{align*}

The 3 operators $\mathcal{R}_{i}[h], i = 1, 2, 3$ can be understood as follows:
\begin{enumerate}
 \item $\mathcal{R}_{1}[h]$ regularizes the integral at the singularity point, by removing 
       a sufficient number of terms from the numerator (2 terms for the power of $5/2$).
       The integration is over a small segment $(\tau'_{0}, \tau'_{h})$.
       
 \item $\mathcal{R}_{2}[h]$ adds the removed terms at the upper boundary $\tau'_{h} = \tau'_{0} + h$.
 
 \item $\mathcal{R}_{3}[h]$  adds the remainder of the integral, which by assumption, is well-defined,  
       since $R(\tau')$ has no more roots in the remaining interval $(\tau'_{h},\tau)$.
\end{enumerate}

%%%Even though $h$ is not particularly defined, it has 2 opposing contraints:
The parameter $h$ should satisfy the following constraints:
\begin{enumerate}
 \item $\dot{R}(\tau') \neq 0$ for $\tau'_{0} < \tau' < \tau'_{h}$. This ensures that $\mathcal{R}_{1}[h]$ and $\mathcal{R}_{2}[h]$ 
       are well-defined.
 \item $h$ cannot be vanishingly small, since that would make the terms contributed by $\mathcal{R}_{2}[h]$ and
       $\mathcal{R}_{3}[h]$ divergent. 
\end{enumerate}

There are cases, depending on the observation point $(x,\tau)$, where the above constraints cannot be met, 
being either mutually exclusive or unattainable altogether
%%However, we would be able to show that these cases correspond exactly to similar conditions in which even
%%Maxwell fields would be divergent.
(these cases correspond to similar conditions in which Maxwell fields would be divergent as well.).

The regularization scheme outlined above, based on the theory of regularization of generalized functions given in Gel'fand 
\cite{Gelfand1964_1}, is used quite frequently in numerical integration of hypersingular integrals
(cf. \cite{zozulya_2006} and references therein).
Indeed, we have used \eqref{def:canonical_regularization_for_case_1} in the numerical evaluation 
of $f^{\alpha \beta}$. 
Similar methods are mentioned in other classical higher dimension electrodynamics applications
\cite{GalTsov2002,Kazinski2002,gurses-2003-44} and and in many examples in quantum field theory.

\section{Conditions for regularizability}
    \label{sec:conditions_for_regularizability}
    \subsection{Behavior at singularities}

As $R(\tau')$ can be understood as a Lorentzian squared distance between the observation point 
$(x,\tau)$ and the source point $(x(\tau'),\tau')$, the derivative 
$\dot{R}(\tau')$ which can be written as
\begin{align}
    \label{eq:r_dot}
    \dot{R}(\tau') & = 2 \dot{z}^{\alpha}(\tau') (x^{\alpha} - z^{\alpha}(\tau'))
\end{align}
is normally interpreted as the \emph{spatial distance} between the observation point and source point 
in the frame where the source is momentarily at rest\footnote{Strictly speaking, it is  $\frac{1}{2}\dot{R}(\tau')$ which would
                                                              be the spatial distance, and not $\dot{R}(\tau')$.}.
However, as events associated with timelike particles are not at rest in the temporal sense nor in the
$\tau$ sense, there is no physical frame in which $\dot{R}(\tau')$ is a pure spatial distance.

In the case of $(3,1)$ Maxwell electrodynamics, $\dot{R}(\tau')$ is the denominator of the 
Li\'{e}nard-Wiechert potentials, which has the well known form
\begin{align}
    \label{eq:lienard_wiechert}
    A^{\mu}(x)
    & = 
        \dfrac{e \dot{z}^{\mu}(s_0)                             }
              {4 \pi \dot{z}_{\mu}(s_0) (x^{\mu} - z^{\mu}(s_0))}
    =
        \dfrac{e \dot{z}^{\mu}(s_0) }
              {2 \pi \dot{R}(s_0)   }
\end{align}
where $s_0$ is defined as the retardation point
\begin{align*}
    R(s_0) & = 0 = - (x_{\mu} - z_{\mu}(s_0))
                     (x^{\mu} - z^{\mu}(s_0))
\end{align*}
with the extra retardation condition $t > z^{0}(s_0)$.

The case of $\dot{R}(\tau') = 0$ is indeed interpreted as where the fields are normally singular,
unless higher order regularization is employed, since they amount to observing the field on the particle itself.
Therefore, in our case as well as in Maxwell electrodynamics, the case of $R(s_{0}) = \dot{R}(s_0) = 0$ leads to $|A^{\mu}(x)| \to \infty$.

As $R(\tau')$ reflects the squared timelike Lorentzian distance, the Green-function support peaks on the light cone
$R(\tau'_{0}) = 0$ for both even and odd spacetime dimensions. Therefore, the combined condition 
\begin{align*}
    R(\tau'_{0}) = \dot{R}(\tau'_{0}) = 0
\end{align*}
indicates that to first order, the wavefront remains on the same observation point, forming a discontinuity in the 
value of the field, similar to a shockwave front.

\subsection{Dependence on source motion}
By defining explicitly the distance 5-vector $\sigma^{\alpha}(\tau') \equiv x^{\alpha} - z^{\alpha}(\tau')$, 
and using some elementary spacetime vector algebra, the following can be shown straightforward:
\begin{itemize}
 \item For a common root $\tau'_{0}$ such that $R(\tau') = \dot{R}(\tau'_{0}) = 0$, 
       we have:
       \begin{itemize}
        \item $\sigma^{\alpha}(\tau'_{0}) \sigma_{\alpha}(\tau'_{0}) = 0$ by definition of $R(\tau'_{0}) = 0$, 
              and thus, $\sigma^{\alpha}(\tau')$  is a \emph{null vector} at $\tau'_{0}$.
        \item $\dot{R}(\tau') = - 2 \dot{z}^{\alpha}(\tau') \sigma_{\alpha}(\tau')$, and thus,
              for $\dot{R}(\tau'_{0}) = 0$, we find that $\dot{z}^{\alpha}(\tau')$ is either \emph{null} or \emph{spacelike} vector
              at $\tau'_{0}$.
        \item If $\dot{z}^{\alpha}(\tau'_{0})$ is a null vector, then it is proportional to $\sigma^{\alpha}(\tau'_{0})$.
              This shows that when the roots are common, the particle is seen to be moving in the (spacetime) direction
              of the observation point, with respect to the source point.
        \item In a Lorentzian local-frame, where the source is momentarily at rest in the \emph{spatial sense}, 
              the spatial distance between observation and source points is \emph{zero}.
       \end{itemize}
       
 \item For a root $\tau''_{0}$ of $\dot{R}(\tau''_{0}) = 0$, which is \emph{not a root of $R(\tau''_{0}) > 0$},
       then $\sigma^{\alpha}(\tau''_{0})$ is a \emph{timelike} vector, whereas $\dot{z}^{\alpha}(\tau''_{0})$ is a 
       \emph{spacelike vector}, as they are orthogonal.
       
 \item If the motion of the source is purely 5D timelike, then there is no $\tau'$ such that $\dot{R}(\tau') = 0$, which 
       means that for this type of motion, the (regularized) fields are \emph{smooth}.
\end{itemize}

\section{Results}
    \label{sec:numerics_and_results}
    %%
%% File   :  UAP_results
%% Purpose:  Showing the numerical results of the simulation
%%
\subsection{On the numerical evaluation of the fields}
Evaluating \eqref{eq:f_field_due_to_point_source} using the generalized regularization, in which a special case is given by
\eqref{def:canonical_regularization_for_case_1} for the fields of a uniformly accelerated source which spacetime trajectory
is given by \eqref{eq:intro_hyperbolic_worldline_1} was done by a numerical computation. 
As the source is moving along the $x$ axis, the $y$ and $z$ origins are chosen such that $y(\tau') = z(\tau') = 0$
for all $\tau'$, and thus, by symmetry, the field has only radial $\rho = (y^2 + z^2)^{1/2}$ dependence in the $yz$ plane.
Only the fields for the case of $(4,1)$ signature were computed.

Therefore, the non-zero $a$ potentials are $a^{\alpha}$ for $\alpha = t, x, \tau$ or $0, 1, 5$, which leaves only 6 
non-zero independent $f$-fields $f^{xt}, f^{x\rho}, f^{x\tau}, f^{t\tau}, f^{t\rho}$ and $f^{\rho\tau}$.
Clearly, $f^{x\rho}$ is normally identified as $B_{\phi}$ in Maxwell electrodynamics
($B_{\phi}$ in this case rotates around the $x$ axis).

The fields were computed on a simple rectangular mesh which spans a rectangular area 
of the $(x,t)$ plane centered around the origin $(x=0,t=0)$, which was repeated for various $\rho$ values.
The fields were \emph{not} evaluated directly on the $xt$ plane itself ($\rho = 0$) as it contains the actual trajectory of the source,
which would be highly singular.
The regularized evaluation of the field on each mesh point is described in \ref{alg:field_computation}.

\begin{algorithm}
    \label{alg:field_computation}
    \caption{Computation of a field $f$ for each mesh point defined by  $x,t,\rho,\tau$}
    \begin{algorithmic}[1]
        \STATE{$\mathcal{R}_{i} = \{ r_{i} \} \leftarrow $ root solution(s) for $\tau'$ of the equation $R(x,t,\rho,\tau,\tau')=0$}
        \STATE{$n_{r} \leftarrow |\mathcal{R}_{i}|$ }
        \STATE {Classify $n_{s} = n_{r}+1$ segments $\{ S_{i} \}$ between roots}
        \\
        \COMMENT{First segment begins with $-\infty$, and last segment ends with $\tau$}
        
            \FOR{$i=0$ to $n_{s}$}
                \IF{$R(\tau') \geq 0$ in segment $S_{i}$}
                    \STATE{$\tau'_{0} \leftarrow $ segment start point}
                    \STATE{$\tau'_{1} \leftarrow $ segment end point  }
                    \IF{$R(\tau'_{0}) = 0$}
                        \STATE{$L_{l} \leftarrow $ regularize-lower-end}
                        \STATE{$\tau'_{0} \leftarrow \tau'_{0} + h$}
                    \ENDIF
                    \IF{$R(\tau'_{1}) = 0$}
                        \STATE{$L_{h} \leftarrow $ regularize-higher-end}
                        \STATE{$\tau'_{1} \leftarrow \tau'_{1} - h$}
                    \ENDIF
                    \STATE{$L \leftarrow $ regular integration from $\tau'_{0}$ to $\tau'_{1}$}
                    \STATE{$I_{i} \leftarrow L + L_{h} + L_{l}$}
                \ENDIF
                \STATE{$I \leftarrow I + L_{i}$}
            \ENDFOR
    \end{algorithmic}
\end{algorithm}

For a given mesh point, the roots of $R(\tau')|_{x}$ were found using high-precision arithmetic library \cite{baily_2002},
as it contains both exponential and quadratic terms in $\tau'$.
The numerical integration was performed using GSL's QAGI, QAGP an QAWS algorithms \cite{gsl_2009}, 
whereas the integrand evaluation was performed using high-precision arithmetics.

Numerical computation plots were made with VisIt (\texttt{www.llnl.gov/visit}), whereas 
analytical plots (in figure \ref{fig:xt_quad_tau_roots}) were made with GNUPLOT (\texttt{www.gnuplot.info}).

We have generally taken $h=0.1$.

\subsection{Notation}
\begin{figure}
 \centering
     \begin{tikzpicture}
        %% 
        %% Grid (gray very thin)
        %% 
        \draw[step=.5cm,color=blue!30,very thin] (-4.5, -4.5) grid (4.5, 4.5);
        %%
        %% Axes lines with arrows (over the grid! not below it)
        %%
        \draw [->,color=gray, thick] (-4.6,0) -- (4.6, 0)  node[right=1pt, color=black] {\Large $x$} coordinate (x axis);
        \draw [->,color=gray, thick] (0, -4.6) -- (0, 4.6) node[above=1pt, color=black] {\Large $t$} coordinate (y axis);
        %%
        %% Draw x=t line.
        %%
        \draw[color=gray!70!black, very thick, dashed]  
             (-4.5, -4.5) -- node[above=2pt, sloped, black] {$x = t$} (-3.0, -3.0) -- (4.5, 4.5) ;
        %%
        %% Draw x=-t line.
        %%
        \draw[-,color=gray!70!black, very thick, dashed]  
             (-4.5,  4.5) -- node[below=2pt, sloped, black] {$x = -t$} (-3.0,  3.0) -- (4.5, -4.5) ;
        %%
        %% Text
        %%
        \draw[color=black] ( 3, 0) node[above=1pt, color=black] {\Large Region I};
        \draw[color=black] ( 0 ,2) node[color=black] {\Large Region II};
        \draw[color=black] (-3, 0) node[above=1pt,  color=black] {\Large Region III};
        \draw[color=black] (0 ,-2) node[color=black] {\Large Region IV};
        %%
        %% Trajectory.
        %%
        \draw[smooth,domain=1:4.5, color=red,samples=100] plot(\x,{  sqrt((\x)*(\x) - 1)  }) ;
        \draw[smooth,domain=1:4.5, color=red,samples=100] plot(\x,{ -sqrt((\x)*(\x) - 1)  }) ;
        %%
        %% Trajectory - parameterized
        %%
        %%\draw[smooth,domain=-2.5:2.5, color=red,samples=100, variable=\t] plot({ exp(\t)/2 + exp(-\t)/2},{exp(\t)/2 - exp(-\t)/2});
        %%
        %% Lines for the trajectory
        %%
        
        %%
        %% Paths along the trajectory
        %% 1st - 2 nodes
        %%
        \coordinate (incomingA) at (2  ,  0.2 + {-sqrt(2  *  2-1)}) ;   
        \coordinate (incomingB) at (1.5,  0.2 + {-sqrt(1.5*1.5-1)}) ;   
        
        \coordinate (outgoingB) at (2  , -0.2 + { sqrt(2  *  2-1)}) ;   
        \coordinate (outgoingA) at (1.5, -0.2 + { sqrt(1.5*1.5-1)}) ;   

        \coordinate (finger)    at (2.5,  0.1 + {-sqrt(2.5*2.5-1)}) ;

        \draw[->, blue, thick] (incomingA)--(incomingB); 
        \draw[->, blue, thick] (outgoingA)--(outgoingB); 

        \draw[<-, black, thick] (finger) -- +(1.0, 1.0) node[above] {Source path};  
        
    \end{tikzpicture}
    \caption{$xt$ light cone structure and region notation convention (color online)}
    \label{fig:source_path}
\end{figure}
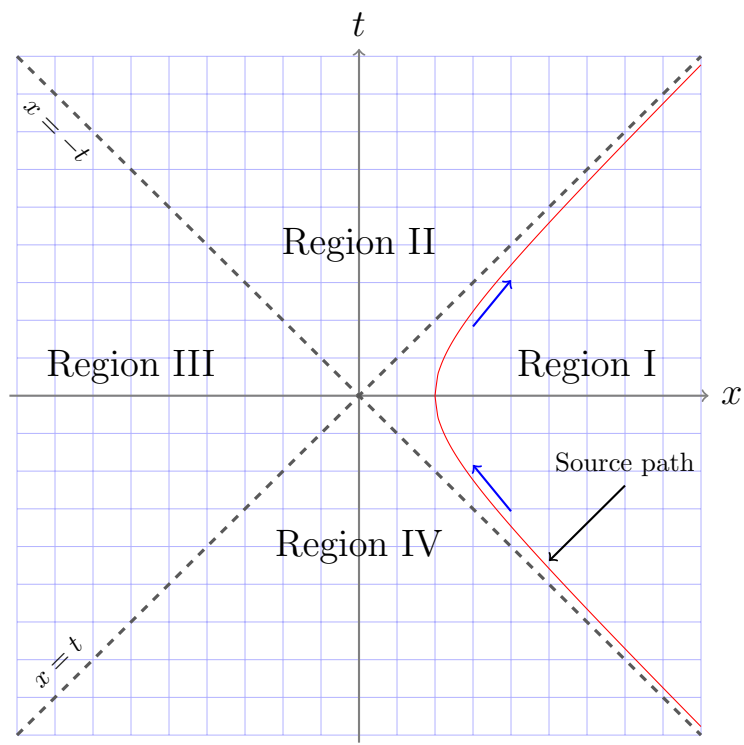

The $(x,t)$ plane is conveniently spanned with a hyperbolic basis, which can (openly) cover only one quarter of the plane.
The division to quadrants and the hyperbolic trajectory follows the convention and is depicted in figure \ref{fig:source_path}.

\subsection{Topography and dynamics of the fields}
Generally, the values of the roots $\tau'_{0}(t,x,\rho,\tau)$ for which $R(\tau'_{0}) = 0$,
determine the common topography of all fields components, as these roots essentially 
designate the intersection of the past light-cone with the vertex at $(t,x,\rho,\tau)$
and the trajectory of the source particle parametrized by $\tau'$
\footnote{E.g., in Phillips's book \cite{lax_2006}, it is shown that most of the contribution to a solution of 
          the wave equation emerges along a characteristic surface, i.e., the null-cone.}

\begin{figure}
 \centering
 \includegraphics[width=15cm]{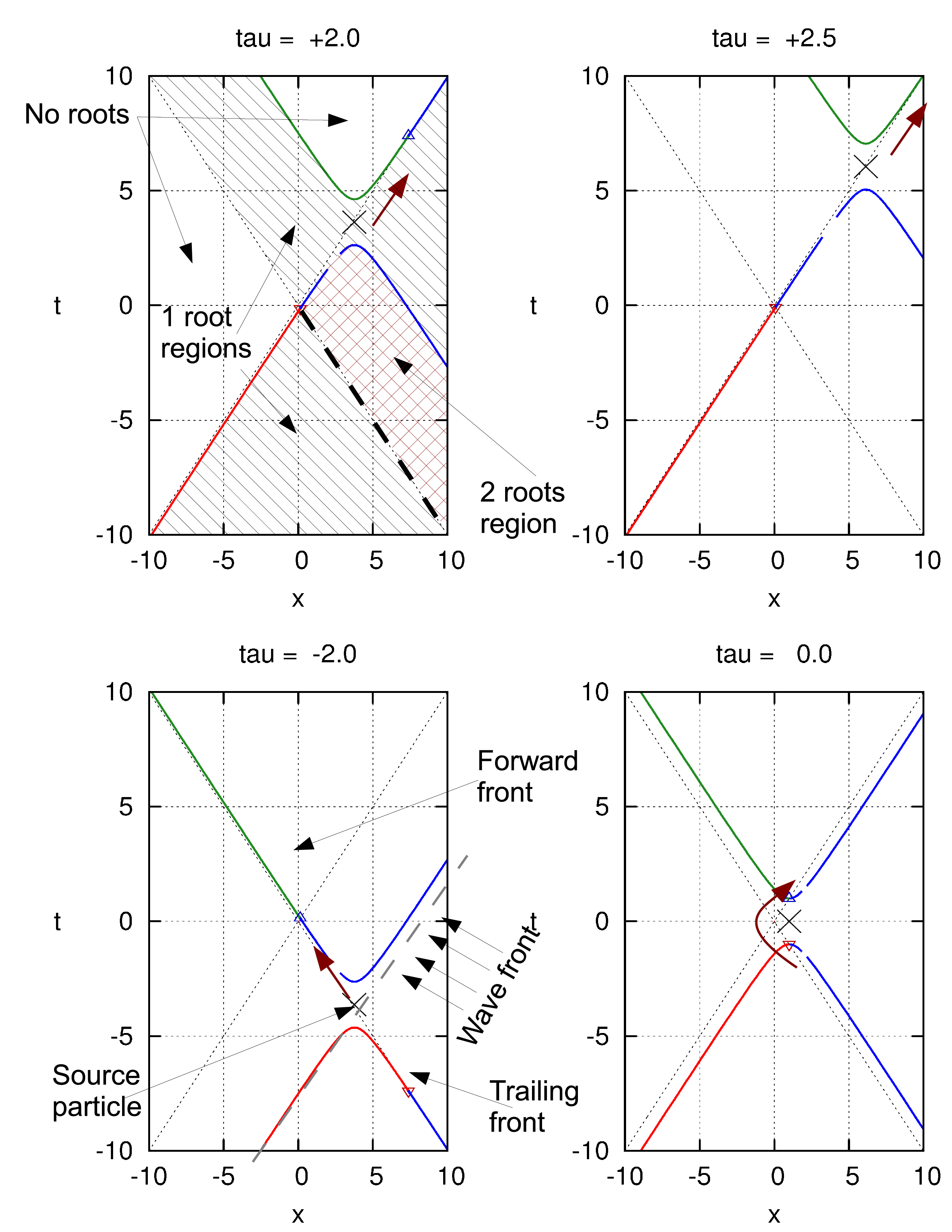}
 \caption[Contour $\tau'_{0}=0$]
         {
             $xt$ location of roots $\tau'_{0}(x,t) = \tau$.
             Each color shows the contour in a different quadrant (color online).
         }
 \label{fig:xt_quad_tau_roots}
\end{figure}

\begin{figure}
 \centering
 \includegraphics[height=13cm]{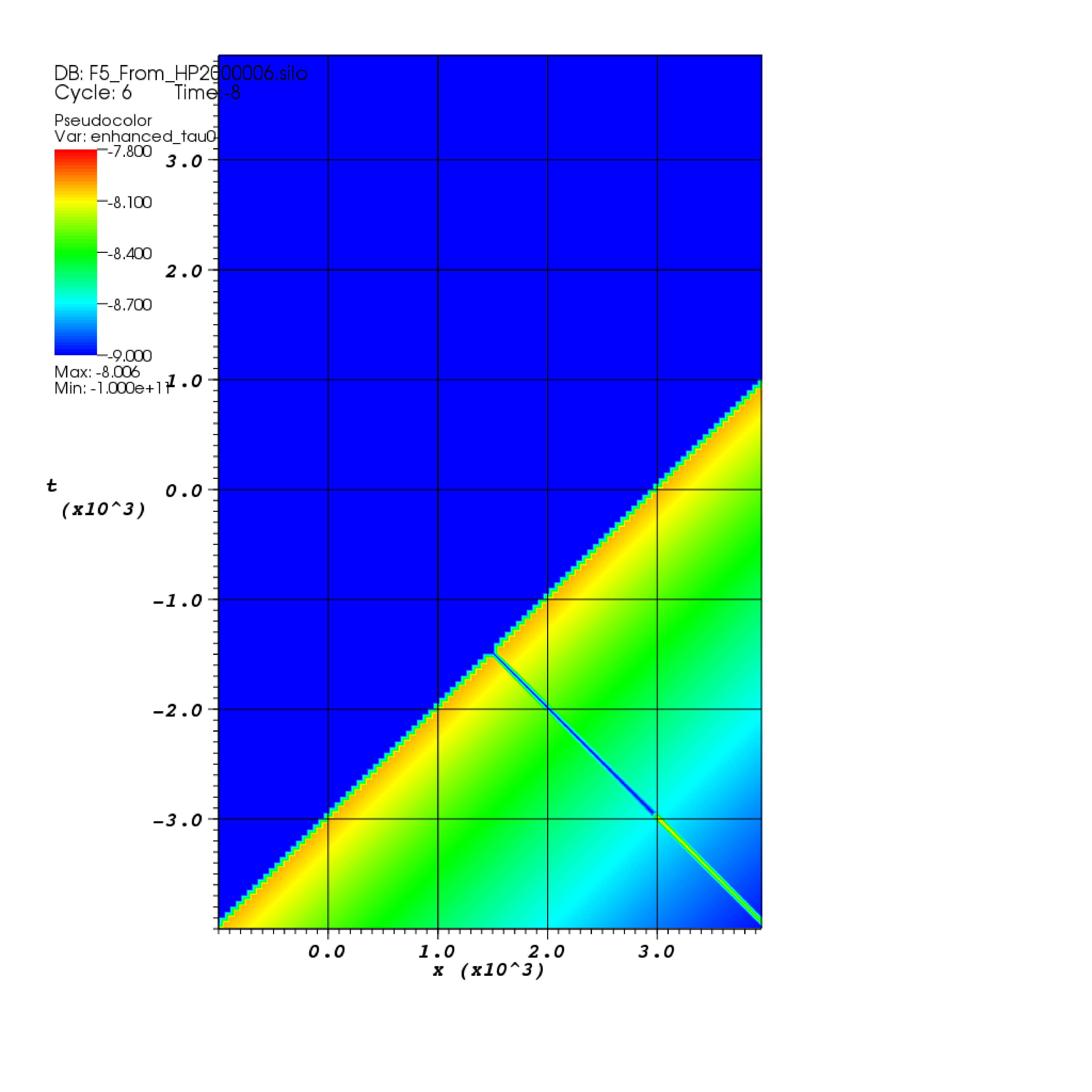}
 \caption[$\tau'_{0}(x,t)$ at $\tau = -8$]
         {Linear plot of the root $\tau'_{0}(x,t)$ obeying $R(\tau'_{0}) = 0$, 
          at $\tau = -8$. Upper part (blue color online) denotes areas where no root exists.
         }
 \label{fig:roots_tau_minus_8}
\end{figure}

\begin{figure}
 \centering
 \includegraphics[height=13cm]{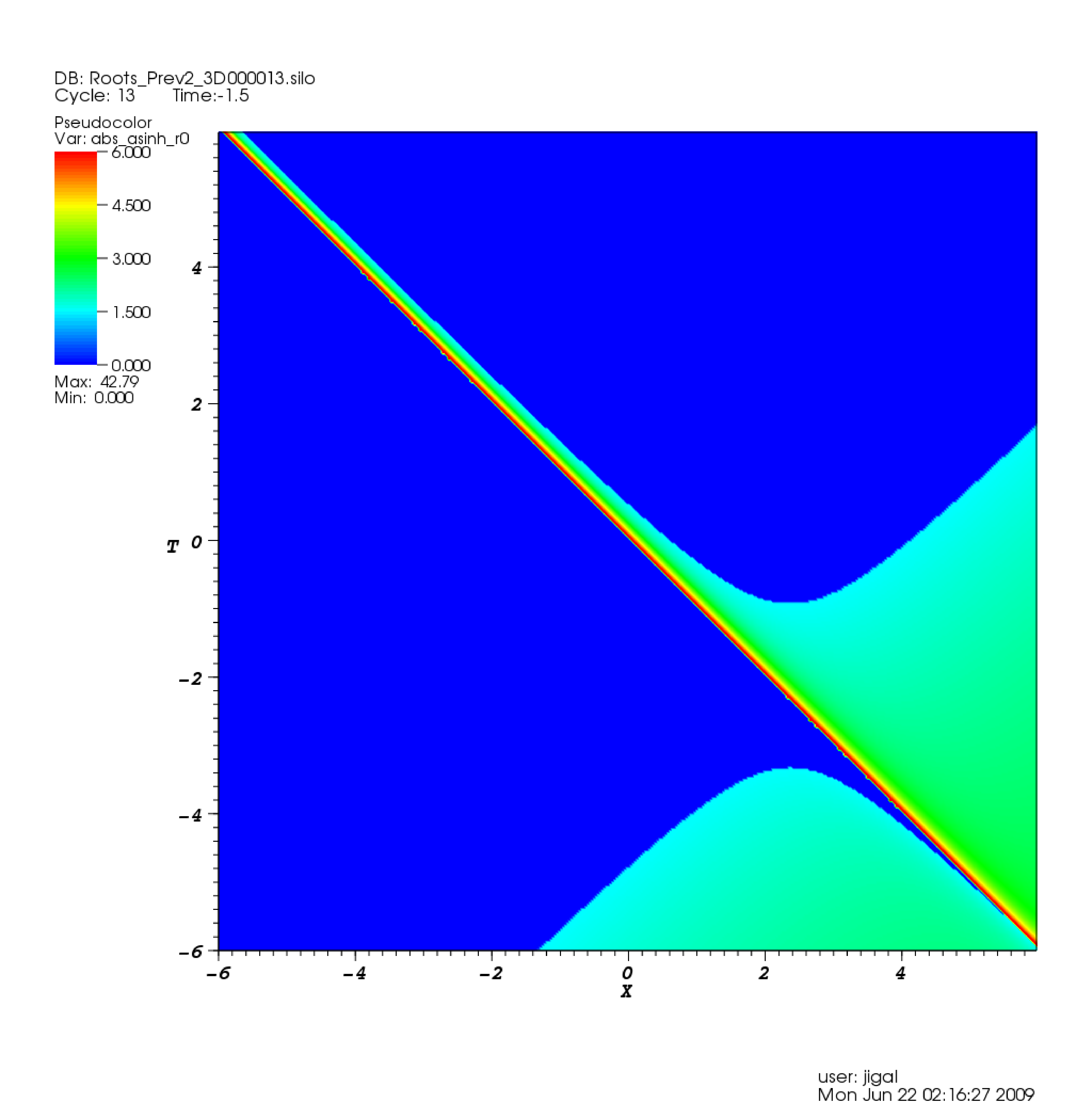}
 \caption[$\tau'_{0}(x,t)$ at $\tau = -1.5$]
         {
            Same linear plot of the root $\tau'_{0}(x,t)$ at much smaller scale, taken at $\tau = -1.5$
            (color online).
         }
 \label{fig:roots_tau_minus_1_5}
\end{figure}

In figure \ref{fig:roots_tau_minus_8} and \ref{fig:roots_tau_minus_1_5}, 
the values of the root are color plotted (online) on the $(x,t)$
plane for $\tau = -8$ and $\tau = -1.5$, respectively.

Even though it is not possible to extract a formal closed-form solution 
$\tau'_{0}(x,t,\rho,\tau)$ such that $R(\tau'_{0}) = 0$, 
one can easily plot \emph{contours} of \emph{iso-root} surface.
In figure \ref{fig:xt_quad_tau_roots} the contour of the 
highest root $\tau'_{0} = \tau$, is shown, for various values of $\tau$,
and it has a special significance:  it is the most
advanced surface observable at a given $\tau$.

Clearly, an iso-root contour is a 2-sided hyperbola which center is the source particle itself at time $\tau$.
This can be shown analytically
as $R(\tau')$ remains form invariant if $x \to x \cosh \alpha + t \sinh \alpha$,
$t \to x \sinh \alpha + t \cosh \alpha$ and $\tau' \to \tau' - \alpha$, which clearly
follows the hyperbolic motion of the source particle.

\begin{figure}
 \centering
 \includegraphics[height=13cm]{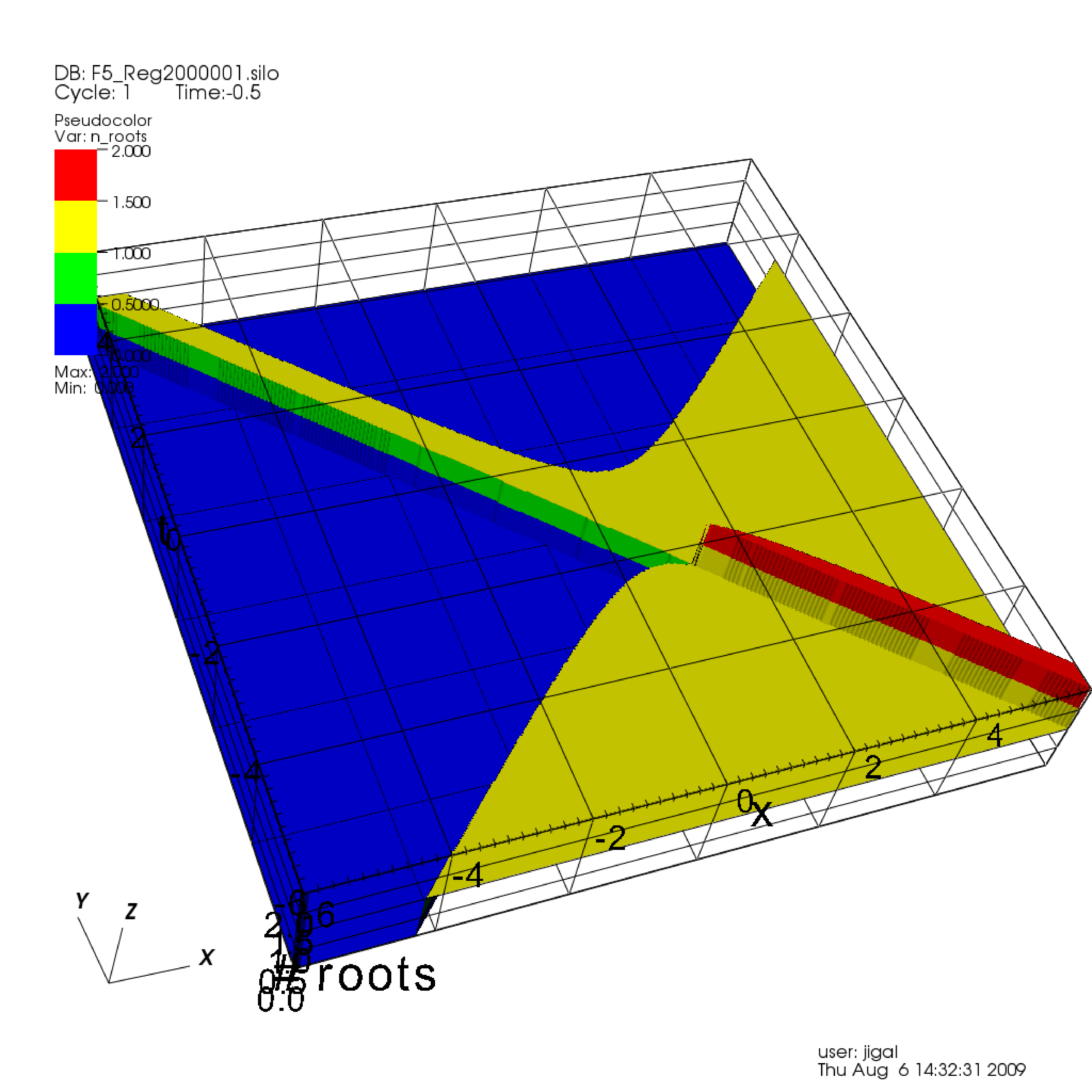}
 \caption{Number of roots at $\tau = -0.5$ (color online)}
 \label{fig:n_roots_tau_minus_0_5}
\end{figure}

Several features can be observed in the iso-root plot,
where we are using figure \ref{fig:xt_quad_tau_roots} as a reference:
\begin{itemize}
 \item The two-sided hyperbola is due to the fact that there are 2 contours with the same root value.
 
 \item The lower side is the \emph{t-advanced} front, at it appears for $t < z^{0}(\tau)$.
       The upper side is then clearly the \emph{t-retarded} front.
 
 \item At $\tau \ll 0$, the lower-side hyperbola forms a \emph{trailing front} behind the particle.
       This is clearly visible in figure \ref{fig:roots_tau_minus_8}.
 
 \item For the same $\tau \ll 0$, the upper-side hyperbola forms an \emph{advanced front}, 
       which approaches the $x=-t$ asymptotically. 
       This advanced front is completely missing in figure \ref{fig:roots_tau_minus_8}, due to 
       the very coarse resolution, whereas it is clearly visible in the finer resolution 
       figure \ref{fig:roots_tau_minus_1_5}.
       
 \item When $\tau \gg 0$, the particle changes its asymptotic direction along the $x=t$ line,
       and the trailing and advanced fronts shift similarly from along the $x=-t$ to the $x=t$ lines.
       
 \item Each contour of the hyperbola has a half that is asymptotically parallel to the particle's motion, 
       and a half that is asymptotically orthogonal to it. The halves are exchanged as the particle 
       crosses the $t=0$ line at $\tau = 0$.
       
 \item For $\tau \ll 0$, the $x=-t$ line behind the particle, starting from the trailing front, 
       is the location of a very high-field value. Similarly, for $\tau \gg 0$ the line  $x=t$ in front
       of the particle, \emph{ending} at the advanced front, is also a location of high field value.
       
       This can be easily seen by inspecting $\dot{R}(\tau'_{0})$ along the $x=\pm t$ lines:
       \begin{align*}
            \dot{R}_{\pm}(\tau') & \equiv \dot{R}(\tau') \Big|_{x = \pm t}
                                   =      2(\tau - \tau') \mp (x \mp t) e^{\mp \tau'}
       \end{align*}
       Clearly, for roots $0 < \tau - \tau'_{0} \ll 1$, we have 
       \begin{align*}
            |\dot{R}_{\pm}(\tau'_{0} \approx \tau)| & \approx 2x e^{- |\tau|} 
       \end{align*}
       Therefore, the fields diverge as $1 / \dot{R}_{\pm}(\tau'_{0})$.
       
 \item Quadrant I is the only quadrant where eventually, every point would covered by two roots of $\tau'_{0}$.
       The region in quadrant I \emph{between the upper and lower sides of the hyperbola} is where a \emph{single root}
       $\tau'_{0}$ exists. 
       The region \emph{below the lower side of the hyperbola} is where two roots exist.
       
       In quadrant IV, the region below the same side of the hyperbola has a single root.
       
       Thus, when crossing the line \emph{$x=-t$ below the lower-side of the hyperbola}, moving from quadrant IV to 
       quadrant I, the number of roots \emph{bifurcates} from $1$ to $2$ roots, 
       which can be seen in figure \ref{fig:n_roots_tau_minus_0_5}.
       
       As $\tau$ progresses, the doubly covered region (red in \ref{fig:n_roots_tau_minus_0_5})
       continues to propagate into quadrant I, trailing the particle.
\end{itemize}

Notably, the $\tau'$ support of the Green-function is different in each quadrant. 
In quadrant IV, the integration takes place from $(\tau'_{0} , \tau)$, 
in II from $(-\infty , \tau'_{0})$ and in I it bifurcates to two ranges, from $(-\infty , {\tau'}_{0}^{(1)})$ 
and $({\tau'}_{0}^{(2)} , \tau)$ where the history ${\tau'}_{0}^{(1)} \leq \tau' \leq {\tau'}_{0}^{(2)}$
remains outside the domain of influence.

For points outside the characteristic $x = \pm t$, the root landscape $\tau'_{0}(x,t,\tau)$ 
changes little when $\tau - \tau'_{0} = h > O(g^{-1})$, and once a point $(x,t)$ is 
inside the past 5D light cone ($\tau'_{0} \leq \tau$), 
its field value remains almost constant once $h = \tau - \tau'_{0}> O(g^{-1})$.

\begin{figure}
 \centering
 \includegraphics[height=13cm]{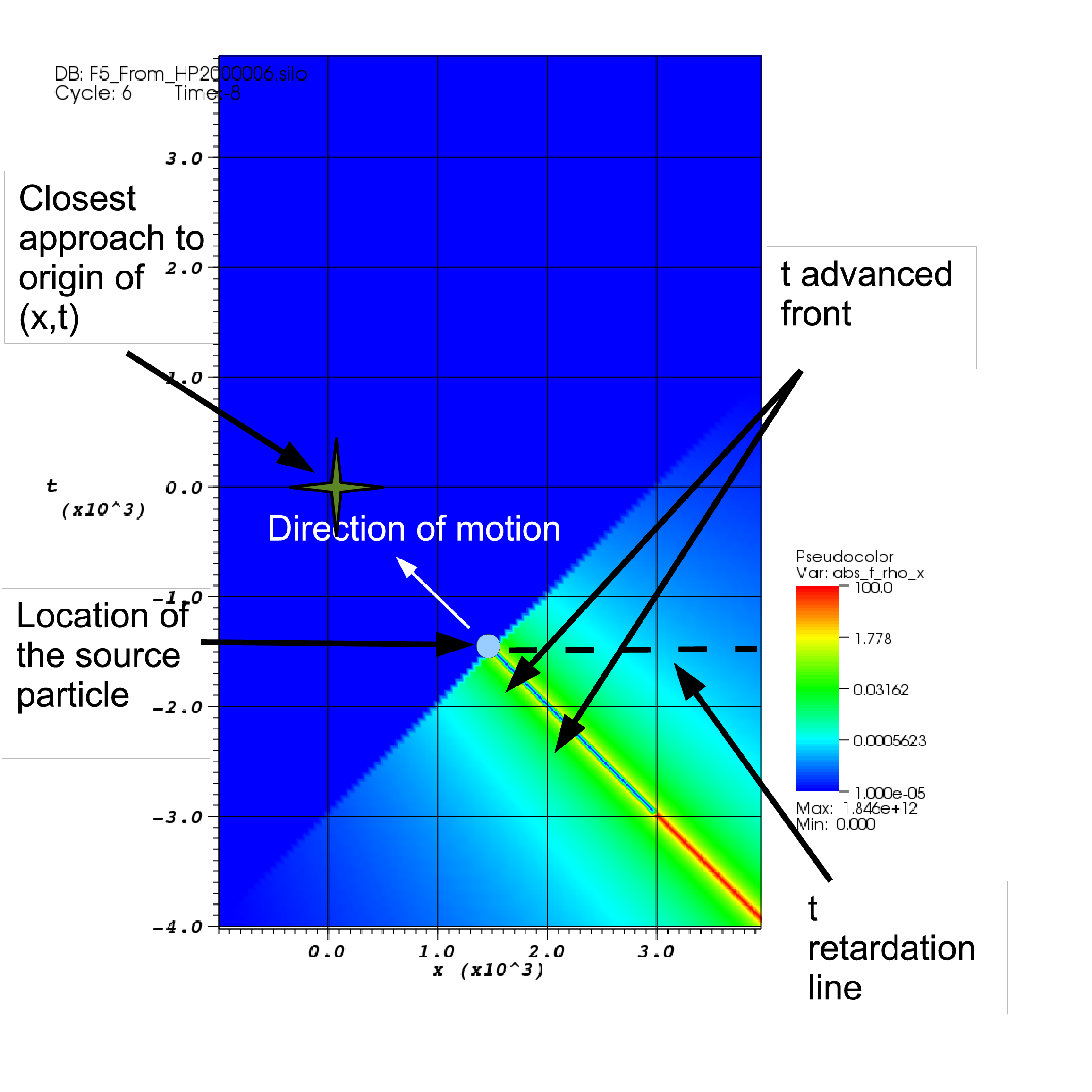}
 % abs_f_x_rho_tau_minus_10.eps: 0x0 pixel, 300dpi, 0.00x0.00 cm, bb=14 14 1039 1144
 \caption[Logarithmic color plot of $|f^{x\rho}|$ at $\tau = -8$, $\rho = 1$]
         {Logarithmic color plot of $|f^{x\rho}|$ at $\tau = -8$, $\rho = 1$.
          The position source particle is shown (color online).
         }
 \label{fig:abs_f_x_rho_tau_minus_8}
\end{figure}

In figure \ref{fig:abs_f_x_rho_tau_minus_8}, a logarithmic color plot of the field $|f^{x\rho}|$ is given 
for $\tau = -8$. The location of the source and the direction of motion are depicted.
Notably, the field is $0$ on the $x=-t$ line up to $x = -t = \frac{1}{2} (\rho^2 + 1) e^{- \tau} \approx 2981$.
Behind the particle there is the trailing front. As the trailing front crosses the $x=-t$ line, 
the fields values there get very high due to  $|\dot{R}_{\pm}(\tau'_{0})| \ll 1$ along these lines.

\begin{figure}
 \centering
 \includegraphics[height=13cm]{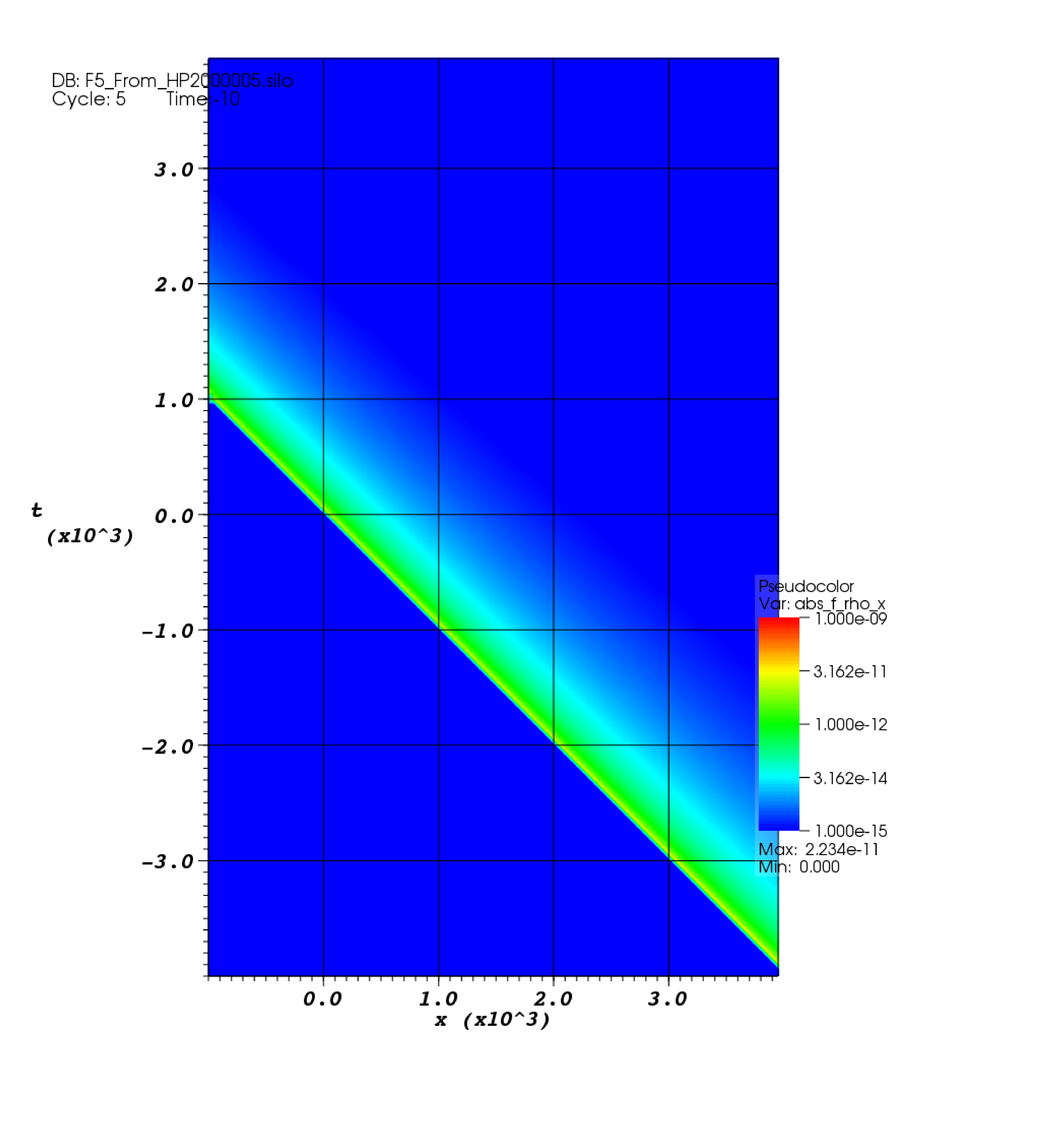}
 % abs_f_x_rho_tau_minus_10.eps: 0x0 pixel, 300dpi, 0.00x0.00 cm, bb=14 14 1039 1144
 \caption[Logarithmic color plot of $|f^{x\rho}|$ at $\tau = -10$, $\rho = 1$]
         {Logarithmic color plot of $|f^{x\rho}|$ at $\tau = -10$, $\rho = 1$. The
          source particle is not visible in this spacetime patch (color online).}
 \label{fig:abs_f_x_rho_tau_minus_10}
\end{figure}

\begin{figure}
 \centering
 \includegraphics[height=13cm]{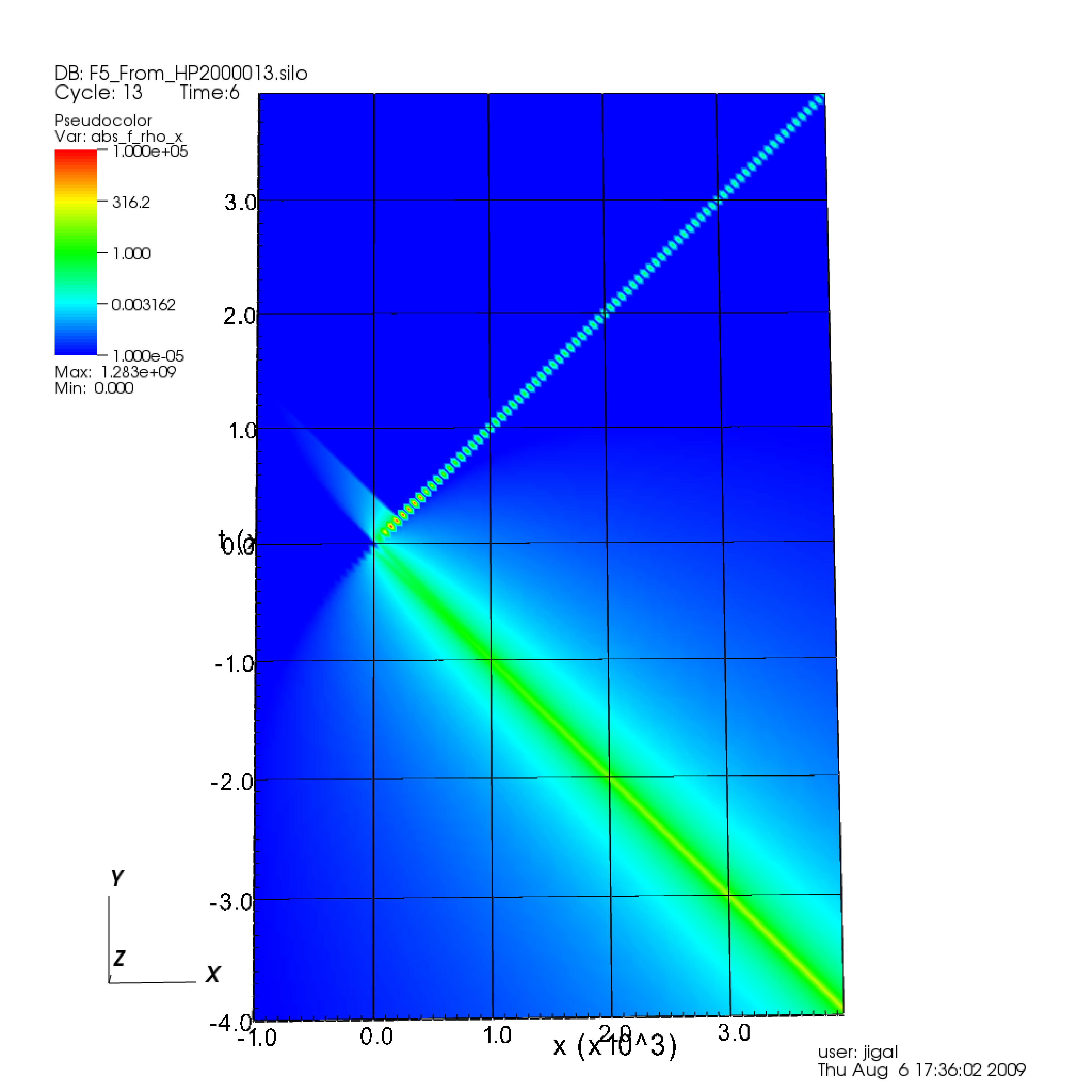}
 \caption[Logarithmic color plot of $|f^{x\rho}|$ at $\tau = 6$]
         {Logarithmic color plot of $|f^{x\rho}|$ at $\tau = 6$
          (color online).
         }
 \label{fig:abs_f_x_rho_tau_minus_6}
\end{figure}

\subsection{Maxwell fields}
\begin{figure}
 \centering
 \includegraphics[height=13cm]{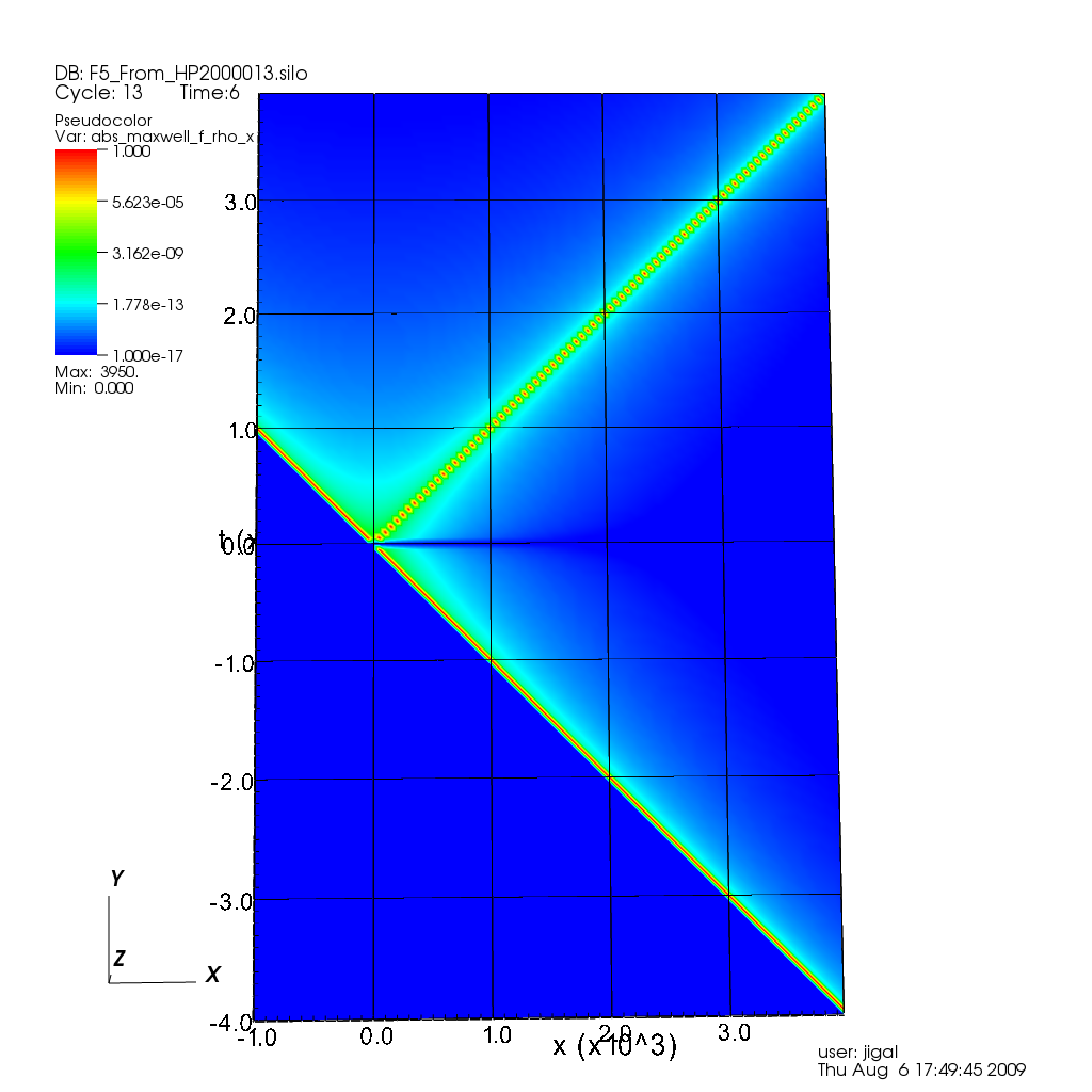}
 \caption[Maxwell $|H_{\phi}| = |f_{\rho x}|$ due to a uniformly accelerated source]
         {Maxwell $|H_{\phi}| = |f_{\rho x}|$ due to a uniformly accelerated source,
          to the same scale as figure \ref{fig:abs_f_x_rho_tau_minus_8}
          (color online)
         }
 \label{fig:abs_maxwell_f_rho_x}
\end{figure}

For comparison, a logarithmic plot of the Maxwell field 
$|H_{\phi}| = |f_{x\rho}|$
as given by \cite{vallisneri_2000}:
\begin{align}
     \label{eq:H_phi_maxwell}
     H_{\phi}
     & = 
         \dfrac{8 e g^{-2} \rho t}
               {
                   \left[
                       \left(
                           g^{-2} - \rho^2 + t^2 - x^2
                       \right)^2
                       + 
                       4 g^{-2} \rho^2
                   \right]^{3/2}
               }
\end{align}
is shown in \ref{fig:abs_maxwell_f_rho_x}, using the same $(x,t)$ plane view and $\rho = 1$ 
distance off the plane of motion (and the same relatively low resolution)

The field shown is evaluated in the region denoted by Einstein retardation $x + t \geq 0$, 
i.e., only quadrants I and II have non-zero field.
The 5D pre-Maxwell offshell fields, on the other hands, exhibit non-zero fields in quadrant IV as well, which are
the source of the \emph{advanced fields} (along with quadrant I).

\subsection{Interpretation}
In the 5D plots of the fields given in figures \ref{fig:abs_f_x_rho_tau_minus_6} ,\ref{fig:abs_f_x_rho_tau_minus_8} 
and \ref{fig:abs_f_x_rho_tau_minus_10},
the fields show a pattern which shares similarity with the Maxwell field as plotted in \ref{fig:abs_maxwell_f_rho_x},
in the sense that both show similar development up the $x=-t$. As the particle decelerates along the $x=-t$ axis, 
the field at the $x=-t$ plane is asymptotically infinite, as it is a buildup of an asymptotically null (in the 4D sense)
particle in the infinite $\tau$-past. 
However, the field builds up on the \emph{negative side} of $x$ as well, even though 
the source particle never visits it.
As the source particle reverses its $x$-velocity at $\tau = 0$, the field begins to buildup along the $x=+t$ surface.

A distinct character of the 5D fields is the $\tau$-motion in the $(x,t)$ plane,
whereas clearly, the Maxwell-Einstein fields, as shown in figure \ref{fig:abs_maxwell_f_rho_x}, 
are essentially \emph{$\tau$-static}.

A \emph{test particle} in the $(x,t)$ plane, would in fact, correlate locally with the 
field at $(x,t)$ at the \emph{same $\tau$}, by coupling to the generalized Lorentz force \eqref{eq:5D_Lorentz_force}.
In this sense, the test particle (or any other type of observer) sees a dynamic spacetime field.
We shall study in a succeeding article, the motion of test particles in the fields that we have investigated.

Furthermore, as seen in figure \ref{fig:xt_quad_tau_roots}, the double sided hyperbola
is in fact strongly related to the fact the Green-functions are \emph{$\tau$-retarded} 
and \emph{not $t$-retarded}, as is the normal case in Maxwell fields\footnote{Or even in higher 
                                                                              dimensional Maxwell electrodynamics,
                                                                              e.g., see \cite{Kazinski2002}, \cite{GalTsov2002}
                                                                                         \cite{gurses-2003-44} and \cite{mironov-2007}.}.
The $\tau$-retarded fields are, in fact, the \emph{average} value of $t$-retarded and $t$-advanced 
fields.
The \emph{lower-side hyperbola} in figure \ref{fig:xt_quad_tau_roots}
is in fact, the wave-front of the \emph{$t$-advanced} field part, 
and the upper-side hyperbola is wave-front of the $t$-retarded part.

\section{Summary and Conclusions}
    \label{sec:summary}
    %%
%% File   :  UAP_summary
%% Purpose:  Summary and conclusions
%%
In order to solve simple problems in 5D off-shell electrodynamics, $\tau$ retarded Green-Functions
(GF's) were necessary, and these were derived using analytic continuation of Nozaki's result 
\cite{Nozaki_1964}, applicable both in $O(4,1)$ and $O(3,2)$ spacetime signature.

The machinery was then applied to a configuration studied long ago in the context
of Maxwell electrodynamics, the radiation of a uniformly accelerated point source,
which has generated decades of debate and though seemingly simple, has elucidated many important
fundamental aspects on the nature of radiation.

However, we note that the vanishing of \emph{radiation reaction} term $\Gamma^{\mu}_{\text{rr}}$
in the equation of motion of the source particle in Maxwell-Einstein electrodynamics 
does not occur in 5D off-shell electrodynamics\footnote{Indeed, the very separation to radiation zone 
                                                       in odd dimensional spacetimes is far 
                                                       less obvious, if it is at all possible.}.

The reason is the support of the GF is on 
the \emph{entire $\tau$ history} of the source particle. Indeed, a full account of the 
off-shell fields in the equation of motion of the source particle is given by
\begin{align}
    M \ddot{x}^{\mu}
    & = 
        q \dot{x}^{\alpha}(\tau) f^{\mu}_{\,\,\, \alpha, \text{ext}}(x(\tau),\tau)
        +
        \nonumber 
    \\
    & \qquad 
        +
        \dfrac{3 \sigma_{5} q^2 }{8 \pi^2}
        \dot{x}^{\alpha}(\tau)
        \mathcal{R}
        \int_{-\infty}^{\tau}
            \dfrac{\theta(R(\tau'))}
                  {R^{5/2}(\tau') }
            \left[
                \dot{x}^{\mu}(\tau')    \partial_{\alpha} R(\tau') 
                - 
                \dot{x}_{\alpha}(\tau') \partial^{\mu} R(\tau')
            \right]
            \ud\tau'
    \label{eq:5d_radiation_reaction_force}
\end{align}
where $R(\tau')$ is identical to \eqref{eq:R_tau_prime}, except that it now
relates the same source particle at different $\tau$ times:
\begin{align}
    R(\tau')
    & = 
        - \sigma_{5} (x(\tau) - x(\tau'))^2
\end{align}
It can easily be shown that the radiation reaction term in \eqref{eq:5d_radiation_reaction_force}
does not vanish for the case of uniform \emph{acceleration}, though it does vanish identically for 
uniform \emph{motion}, which suggests that there is radiation reaction force in the accelerating case.

Generally, the $\tau$-retarded fields can \emph{further} be decomposed to the sum of $t$-retarded and $t$-advanced fields.
Locally, a test particle interacts with both these fields.

\appendix
\section{Canonical regularization of divergent integrals}
    \label{sec:canonical_regularization}
    %%
%% File   : UAP_appendix_canonical_regularization.tex
%% Purpose: Regularization ala Gel'fand
%%
In this section we provide a short overview of the regularization method described in 
Gel'fand \cite{Gelfand1964_1}.

The function $x^{\lambda}_{+} \equiv \theta(x) x^{\lambda}$ is non-zero for positive $x$, where
$\theta(x)$ is the step-function.
When acting on a smooth bounded function $\phi(x)$ 
\begin{align*}
    \left( x_{+}^{\lambda} , \phi(x) \right) 
    & = 
        \int_{0}^{\infty}  x^{\lambda} \phi(x) \, \ud x
\end{align*}
is well defined for $\Re \lambda > -1$. 
On the other hand, the expression can be rewritten as
\begin{align}
    \label{eq:x_lambda_plus}
    \left( x_{+}^{\lambda} , \phi(x) \right) 
    & = 
        \int_{0}^{b}
            x^{\lambda}
            \left[
                \phi(x) 
                -
                \sum_{j=0}^{m}
                    \dfrac{\phi^{(j)}(0)}
                          {j!           }
                    x^{j}
            \right]
            \ud x
        +
        \sum_{j=0}^{m}
            \dfrac{\phi^{(j)}(0)}{j! (\lambda + j + 1)} b^{\lambda + j + 1}
        +
        \nonumber
    \\
    & \qquad
        +
        \int_{b}^{\infty}  x^{\lambda} \phi(x) \, \ud x
\end{align}
where the right-hand-side is well defined for $\{ \Re \lambda > -m \} \cap \{ \lambda \neq -1, -2,  \}$.

This suggests that, as a \emph{generalized function}, 
$x_{+}^{\lambda}$ can be \emph{defined} by its action on any smooth bounded function $\phi(x)$,
as given by \eqref{eq:x_lambda_plus}. The result is a function of $\lambda$ defined for 
all $\Re \lambda > -m$ except at $\lambda = -1, -2, \ldots -m+1$ where it has \emph{simple poles}
with residues $\dfrac{\phi^{(j)}(0)}{j!}$. This suggests that $x_{+}^{\lambda}$ itself 
is a generalized function with simple poles given by
\begin{align*}
    \text{Res } x_{+}^{\lambda} \Big|_{\lambda = -n}
    & = 
        (-1)^{n}\dfrac{\delta^{(n)}(x)}{n!}
\end{align*}

Similarly, given 2 smooth functions, $\phi(x)$ and $R(x)$, we are seeking 
a regularized solution for
\begin{align}
    \label{eq:R_lambda_definition}
    \left( R_{+}^{-\lambda}(x) , \phi(x) \right)
    & = 
        \int_{a}^{b} \dfrac{\phi(x)}{R^{\lambda}(x)}  \ud x
\end{align}
where, $a$ is defined by $R(a) = 0$, and $R(x) > 0$ for $x \in (a,b)$\footnote{In the meantime, we assume $R(b) > 0$.}.
One can select $c$ such that $a < c < b$ and $\dot{R}(x) \equiv dR/dx \neq 0$ for all $x \in [a,c]$.
Setting $h = c-a > 0$, we find
\begin{align}
    \label{eq:R_lambda_definition_1}
    \left( R_{+}^{-\lambda}(x) , \phi(x) \right)
    & = 
        \int_{a}^{a+h} \dfrac{\phi(x)}{R^{\lambda}(x)}  \ud x
        +
        \int_{a+h}^{b} \dfrac{\phi(x)}{R^{\lambda}(x)}  \ud x
    \\
    & = 
        \int_{0}^{R_{h}} \dfrac{\phi(x(R))}
                               {R^{\lambda}} 
                         \dfrac{dR}{\dot{R}(x(R))}
        +
        \int_{a+h}^{b} \dfrac{\phi(x)}{R^{\lambda}(x)}  \ud x
\end{align}
where the first integral in $x$ was transformed to an integral in $R$, since $dR > 0$ 
in the interval $(a,a+h)$.
One can then proceed with the regularization as given in \eqref{eq:x_lambda_plus}
\begin{align*}
%%    \label{eq:R_lambda_definition_1}
    \left( R_{+}^{-\lambda}(x) , \phi(x) \right)
    & = 
        \int_{0}^{R_{h}}
            R^{-\lambda}
            \left[
                \dfrac{\phi(R)}{\dot{R}(R)}
                -
                \sum_{j=0}^{m}
                    \dfrac{R^{j}}{j!}
                    \dfrac{\ud^{j}}{\ud R^{j}}  \left( \dfrac{\phi(a)}{\dot{R}(a)} \right)
            \right]
            \ud R
        +
    \\
    & \qquad 
        +
        \sum_{j=0}^{m}
            \dfrac{R^{j - \lambda + 1}_{h}}{(j - \lambda + 1) j!}
            \dfrac{\ud^{j}}{\ud R^{j}}  \left( \dfrac{\phi(a)}{\dot{R}(a)} \right)
        +
        \int_{a+h}^{b} \dfrac{\phi(x)}{R^{\lambda}(x)}  \ud x
\end{align*}
where $R_{h} = R(a+h)$. Once the regularized integral is given in $R$, it can be transformed back to $x$:
\begin{align*}
    \dfrac{d}{dR}
    & = 
        \dfrac{\ud x}{\ud R} \dfrac{\ud}{\ud x}
    =
        \dfrac{1}{\dot{R}(x)} \dfrac{\ud}{\ud x}
    \intertext{Thus:}
    \left( R_{+}^{-\lambda}(x) , \phi(x) \right)
    & = 
        \int_{a}^{a+h}
            R^{-\lambda}(x)
            \left[
                \phi(x)
                -
                \sum_{j=0}^{m}
                    \dfrac{R^{j}(x)}{j!}
                    \left(
                        \dfrac{1}{\dot{R}(x)}
                        \dfrac{d}{dx}
                    \right)^{j}
                    \left( \dfrac{\phi(a)}{\dot{R}(a)} \right)
            \right]
            \ud x
        +
    \\
    & \qquad 
        +
        \sum_{j=0}^{m}
            \dfrac{R^{j - \lambda + 1}_{h}}{(j - \lambda + 1) j!}
            \left(
                \dfrac{1}{\dot{R}(x)}
                \dfrac{\ud}{\ud x}
            \right)^{j}
            \left( \dfrac{\phi(a)}{\dot{R}(a)} \right)
        +
        \int_{a+h}^{b} \dfrac{\phi(x)}{R^{\lambda}(x)}  \ud x
\end{align*}
from which the regularization \eqref{def:canonical_regularization_for_case_1} 
can readily be obtained by setting $\lambda = 5/2$ and $m=1$.

\section{Derivation of the $\tau$ retarded Green Function}
    \label{sec:derivation_of_retarded_green_function}
     %%
%% File   : UAP_appendix_retarded_green_function.tex
%% Purpose: Derivation of the tau retarded green-function
%%
The derivation of $\tau$ retarded Green-functions was made with essentially 
the same method as Y. Nozaki \cite{Nozaki_1964}, which defined a generalized
Riemann-Liouville type integrodifferential operator for ultrahyperbolic spaces,
itself a generalization of an earlier monumental work by M. Riesz
\cite{riesz_1938,riesz_1949}.

Following Nozaki \cite{Nozaki_1964}, we shall show $L^{\alpha}$
can generate the fundamental solution of the $\Delta_{p,q}$ wave-operator
with retardation in $x_{m}$. 

There are three stages involved in order to establish 
the desired Green-Function is indeed the kernel of the operator $L^{\alpha}$.
This is the sibject of the next paragraphs.

\subsection*{Normalization constant $N_{m}(\alpha)$}
The first step is to establish $L^{\alpha}$ by finding $N_{m}(\alpha)$,
the normalization constant, such that 
\begin{align}
    \label{eq:L_alpha_on_exp_x_m}
    (L^{\alpha} e^{x_{m}})(x) & = e^{x_{m}}
\end{align}
This is akin to the Riemann-Liouville operator in one dimension
\begin{align*}
    (I^{\alpha} e^{x})(x) & = e^{x}
                            = \dfrac{1}{\Gamma(\alpha)}
                              \int_{-\infty}^{x}
                                  (x-y)^{\alpha - 1}
                                  e^{y}
                                  \ud y
\end{align*}
which can easily be proven by changing variables to $u = x-y$.

Thus, we have:
\begin{align*}
    (L^{\alpha} e^{x_{m}})(x) & = 
    \dfrac{1}{N_{m}(\alpha)}
    \int_{\mathcal{D}_{x}}  
        r^{\alpha - m}(x - y)
        e^{y_{m}}
        \ud^{m} y
\end{align*}
Changing variables $y_{i} \to x_{i}-y_{i}$ we obtain:
\begin{align*}
    (L^{\alpha} e^{x_{m}})(x) & = 
    \dfrac{1}{N_{m}(\alpha)}
    \int_{\mathcal{D}_{0}}  
        r^{\alpha - m}(y)
        e^{x_{m} - y_{m}}
        \ud^{m} y
    =
    \dfrac{e^{x_{m}}}{N_{m}(\alpha)}
    \int_{\mathcal{D}_{0}}  
        r^{\alpha - m}(y)
        e^{- y_{m}}
        \ud^{m} y
\end{align*}
Making the substitution
\begin{align*}
    s^2 & = \sum_{i=1}^{p} y_{i}^2 ,    
    \qquad
    t^2   = \sum_{j=1}^{q} y_{p+j}^2,
    \qquad
    r^{\alpha - m}(y)
          = (s^2 - t^2)^{(\alpha - m)/2}
    \\
    \ud^{m} y
    & = 
        s^{p-1} t^{q-1} \ud s \ud t \ud\Omega_{p} \ud\Omega_{q},
    \qquad
    y_{m}
          = t \cos(\theta_{q-1})
\end{align*}
we find:
\begin{align*}
    (L^{\alpha} e^{x_{m}})(x) & = 
    \dfrac{e^{x_{m}}}{N_{m}(\alpha)}
    \int_{0}^{\infty}   t^{q-1} \ud q
    \int_{t}^{\infty}   s^{p-1} \ud s
    \int_{S^{p  }}  \ud\Omega_{p  }
    \int_{S^{q-1}}  \ud\Omega_{q-1}
    \\
    & \qquad 
        \times
        \int_{0}^{\pi/2} \sin^{q-2} (\theta_{q-1}) \ud \theta_{q-1}
            (s^2 - t^2)^{(\alpha - m)/2}
            e^{- t \cos(\theta_{q-1})}
\end{align*}
where $\theta_{q-1}$ is integrated from $0$ to $\pi/2$ signifying retardation in $x_{m}$.
We begin with the integration over $s$, by changing variables to $w(s) = t^2/s^2$ which leads to
%%
%% w(s) = t^2 / s^2 
%% dw(s) = - 2 t^2 / s^3 ds = -  2 t^{-1} t^3/s^3 ds = - 2 t^{-1} w \sqrt{w} 
%% ds(w) = - 1/2 t w^{-3/2} dw 
%% s^{p-1} = (s^2)^((p-1)/2) = (s^2/t^2)^((p-1)/2) . t^(+ 2 (p-1)/2 ) = w^(-(p-1)/2) . t^(p-1)
%% (s^2 - t^2) = s^2 ( 1 - t^2/s^2) = t^2 w^{-1} (1 - w)
%% w(s=\infty) = 0
%% w(s=t)      = 1
%% 
%%

\begin{align*}
    \int_{t}^{\infty}   s^{p-1} & 
        (s^2 - t^2)^{(\alpha - m)/2}
        \ud s
    =
    \\
    & = 
        \dfrac{1}{2}
        t^{1 + 2(p-1)/2 } t^{2(\alpha - m)/2} 
        \int_{0}^{1}   w^{-(p-1)/2 - (\alpha - m)/2 + 3/2} 
            (1 - w)^{(\alpha - m)/2} 
            \ud w
    \\
    & = 
        \dfrac{1}{2}
        t^{\alpha - m+p} 
        \int_{0}^{1}   
            w^{(m - p - \alpha)/2 - 1} 
            (1 - w)^{(\alpha - m)/2+1 - 1} 
            \ud w
    \\
    & = 
        t^{\alpha - m + p} 
        B \left( \dfrac{m - p - \alpha}{2} , \dfrac{\alpha - m}{2} + 1\right)
        \qquad \qquad
        \text{where $B(x,y)$ is the \emph{Beta function}.}
    \\
    & = 
        t^{\alpha - q} 
        \dfrac
        %%% numerator   %%%
        {
            \Gamma
                \left(
                    \dfrac{q - \alpha}{2}
                \right)
            \Gamma
                \left(
                    \dfrac{\alpha- m}{2} 
                    +
                    1
                \right)
        }
        %%% denominator %%%
        {
            \Gamma
                \left(
                    1
                    -
                    \dfrac{p}{2} 
                \right)
        }
\end{align*}
Thus, $L^{\alpha} e^{x_{m}}$ takes the form
\begin{align*}
    (L^{\alpha} e^{x_{m}})(x) 
    & = 
        \dfrac{e^{x_{m}}}{N_{m}(\alpha)}
        \dfrac
        %%% numerator   %%%
        {
            \Gamma
                \left(
                    \dfrac{q - \alpha}{2}
                \right)
            \Gamma
                \left(
                    \dfrac{\alpha- m}{2} 
                    +
                    1
                \right)
        }
        %%% denominator %%%
        {
            \Gamma
                \left(
                    1
                    -
                    \dfrac{p}{2} 
                \right)
        }
        \int_{S^{p}  } d\Omega_{p} 
        \int_{S^{q-1}} d\Omega_{q-1}
    \\
    & \qquad 
        \times
        \int_{0}^{\pi/2}
            \sin^{q-2}(\theta_{q-1}) \ud \theta_{q-1}
        \int_{0}^{\infty} t^{q-1} 
            t^{\alpha - q}
            e^{-t \cos(\theta_{q-1})}
            \ud t
\end{align*}

Integration over $S^{p}$ has the usual result $2\pi^{p/2} / \Gamma(p/2)$, and a similar result for the integration
over $S^{q-1}$. The $t$ integration is performed by making the substitution $w = t \cos(\theta_{q-1})$
(and carefully noting that $\cos \theta_{q-1} \geq 0$ in the range of integration):
%%
%% w(t)   = t \cos(\theta_{q-1}) 
%% dw(t)  = \cos(\theta_{q-1})  dt 
%% dt(w)  = dw / \cos(\theta_{q-1})
%% w(t=0) = 0
%% w(t=infty) = \infty
%%
\begin{align*}
    \int_{0}^{\infty} t^{q-1} 
        &
        t^{\alpha - q}
        e^{-t \cos(\theta_{q-1})}
        \ud t
    =
        \int_{0}^{\infty} t^{\alpha-1} 
            e^{-t \cos(\theta_{q-1})}
            \ud t
    \\
    & = 
        \dfrac{1}{\cos(\theta_{q-1})}
        \int_{0}^{\infty}
            \left(
                \dfrac{w}{\cos(\theta_{q-1})}
            \right)^{\alpha - 1}
            e^{-w}
            \ud w
    \\
    & =
        \dfrac{1}{\cos^{1 + \alpha - 1}(\theta_{q-1})}
        \int_{0}^{\infty}
            w^{\alpha - 1} e^{-w}
            \ud w
    \\
    & = 
        \dfrac{\Gamma(\alpha)}{\cos^{1 + \alpha - 1}(\theta_{q-1})}
\end{align*}

Thus:
\begin{align*}
    (L^{\alpha} e^{x_{m}})(x) 
    & = 
        \dfrac{e^{x_{m}}}{N_{m}(\alpha)}
        \dfrac
        %%% numerator   %%%
        {
            \Gamma
                \left(
                    \dfrac{q - \alpha}{2}
                \right)
            \Gamma
                \left(
                    \dfrac{\alpha- m}{2} 
                    +
                    1
                \right)
        }
        %%% denominator %%%
        {
            \Gamma
                \left(
                    1
                    -
                    \dfrac{p}{2} 
                \right)
        }
        \dfrac{2 \pi^{p    /2}}{\Gamma(p    /2)}
        \dfrac{2 \pi^{(q-1)/2}}{\Gamma((q-1)/2)}
        \Gamma(\alpha)
    \\
    & \qquad 
        \times
        \int_{0}^{\pi/2}
            \dfrac{\sin^{q-2}(\theta_{q-1}) }{\cos^{\alpha}(\theta_{q-1})}
            \ud \theta_{q-1}
\end{align*}
where the $\theta_{q-1}$ integration can be taken by using the identity
\begin{align}
    B(a,b)
    & = 
        2 \int_{0}^{\pi/2}
            \sin^{2a-1}(\theta)
            \cos^{2b-1}(\theta)
            \ud \theta
\end{align}

Thus, one finally has 
\begin{align*}
    (L^{\alpha} e^{x_{m}})(x) 
    & = 
        \dfrac{e^{x_{m}}}{N_{m}(\alpha)}
        \dfrac
        %%% numerator   %%%
        {
            \Gamma
                \left(
                    \dfrac{q - \alpha}{2}
                \right)
            \Gamma
                \left(
                    \dfrac{\alpha- m}{2} 
                    +
                    1
                \right)
        }
        %%% denominator %%%
        {
            \Gamma
                \left(
                    1
                    -
                    \dfrac{p}{2} 
                \right)
        }
        \dfrac{2 \pi^{p    /2}}{\Gamma(p    /2)}
        \dfrac{2 \pi^{(q-1)/2}}{\Gamma((q-1)/2)}
        \Gamma(\alpha)
    \\
    & \qquad 
        \times
        \dfrac{1}{2}
        \dfrac
        %%% numerator %%%%
        {
            \Gamma
                \left(
                    \dfrac{q-1}{2}
                \right)
            \Gamma
                \left(
                    \dfrac{1 - \alpha}{2}
                \right)
        }
        %%% denominator %%%%
        {
            \Gamma
                \left(
                    \dfrac{q - \alpha}{2}
                \right)
        }
\end{align*}
And since $L^{\alpha} e^{x_{m}} = e^{x_{m}}$ we find:
\begin{align}
    \label{eq:N_m_alpha}
    N_{m}(\alpha)
    & = 
        2 \pi^{(m-3)/2}
        \Gamma(\alpha)
        \sin \left( \dfrac{p \pi}{2} \right)
        \Gamma \left( 1 - \dfrac{m - \alpha}{2}\right)
        \Gamma \left( 1 - \dfrac{\alpha}{2}     \right)
\end{align}
which clearly coincides with \eqref{eq:jigal_normalization}.

\subsection*{Operation under the ultrahyperbolic d'Alembert operator}
Writing
\begin{align*}
     L^{\alpha}f 
     & = 
         \int_{\mathcal{D'}_{x}} \Phi_{\alpha}(x-y) f(y) \ud^{m} y
\end{align*}
where $\Phi_{\alpha}(x) = r^{\alpha - m}(x) / N_{m}(\alpha)$, we can evaluate 
the action of $\Delta \Phi_{\alpha}(x)$ directly:
\begin{align*}
    \Delta \Phi_{\alpha}(x)
    & = 
        \dfrac{1}{N_{m}(\alpha)} \Delta r^{\alpha - m}(x)
    = 
        \dfrac{1}{N_{m}(\alpha)} (\alpha - m)(\alpha - 2) r^{\alpha - m-2}(x)
    \\
    & = 
        \dfrac{N_{m}(\alpha - 2)}{N_{m}(\alpha)} (\alpha - m)(\alpha - 2) \Phi_{\alpha-2}(x)
\end{align*}
Now:
\begin{align*}
    \dfrac{N_{m}(\alpha - 2)}{N_{m}(\alpha)} 
    & = 
        \dfrac
        %%% numerator %%%
        {
            \Gamma
                \left(
                    \dfrac{\alpha - m}{2}
                \right)
            \Gamma
                \left(
                    \dfrac{3 - \alpha}{2}
                \right)
            \Gamma
                \left(
                    \alpha - 2
                \right)
        }
        %%% denominator %%%
        {
            \Gamma
                \left(
                    \dfrac{2 + \alpha - m}{2}
                \right)
            \Gamma
                \left(
                    \dfrac{1 - \alpha}{2}
                \right)
            \Gamma
                \left(
                    \alpha
                \right)
        }
    \\
    & = 
        \dfrac
        %%% numerator %%%
        {
            \Gamma
                \left(
                    \dfrac{\alpha - m}{2}
                \right)
            \left[
                \left(
                    \dfrac{1 - \alpha}{2}
                \right)
                \Gamma
                    \left(
                        \dfrac{1 - \alpha}{2}
                    \right)
            \right]
            \Gamma
                \left(
                    \alpha - 2
                \right)
        }
        %%% denominator %%%
        {
            \left[
                \left(
                    \dfrac{\alpha - m}{2}
                \right)
                \Gamma
                    \left(
                        \dfrac{\alpha - m}{2}
                    \right)
            \right]
            \Gamma
                \left(
                    \dfrac{1 - \alpha}{2}
                \right)
            \Big[
                (\alpha - 2) (\alpha - 1)
                \Gamma
                    \left(
                        \alpha - 2 
                    \right)
            \Big]
        }
    \\
    & = 
        -
        \dfrac{1}{(\alpha - m)(\alpha - 2)}
\end{align*}
which immediately leads to
\begin{align}
    \Delta \Phi_{\alpha}(x)
    & = 
        - \Phi_{\alpha - 2}(x)
    \Longrightarrow
        \Delta (L^{\alpha}f)(x)
    =
        (L^{\alpha-2}f)(x)
\end{align}

\subsection*{$L^{\alpha}$ near $\alpha \to 0$}

In \cite{Nozaki_1964}, it was shown that 
\begin{align}
    \lim\limits_{\alpha \to 0^{+}} (J^{\alpha} f)(x)
    & = 
        f(x) 
    \Longrightarrow
        \lim\limits_{\alpha \to 0^{+}} J^{\alpha} 
    =
        \delta(x)
\end{align}
which was shown by extending the operation of $J^{\alpha}$ to include its \emph{advanced counterpart} $\bar{J}^{\alpha}$.
The same exact proof can also be established here by extending $L^{\alpha}$, which essentially
produces the same operator $L^{\alpha} + \bar{L}^{\alpha} = J^{\alpha} + \bar{J}^{\alpha}$.

Therefore, one immediately has 
\begin{align}
    \lim\limits_{\alpha \to 0^{+}} L^{\alpha} & = \delta(x)
\end{align}

This means that the Green-Function is essentially $-\Phi_{2}(x)$, as 
\begin{align}
    \Delta \Phi_{2}(x) & = - \Phi_{0}(x) = - \delta(x)
\end{align}
as originally desired, leading to \eqref{eq:jigal_ultrahyperbolic_green_functions}.

\bibliography{bibliography-all}
\bibliographystyle{hplain}
\end{document}